\documentclass[final,5p,times,twocolumn]{elsarticle}

\usepackage{amssymb}

\usepackage{textgreek}
\usepackage{amsmath}
\usepackage{tabularx, multirow}
\usepackage{float}

\journal{Nuclear Instruments and Methods in Physics Research 
Section B: Beam Interactions with Materials and Atoms}

\begin{document}

\begin{frontmatter}



\title{Particle-in-Cell Techniques for the Study of Space Charge Effects
in the Advanced Cryogenic Gas Stopper}


\author[FRIB]{R.~Ringle\corref{cor1}}
\ead{ringle@frib.msu.edu}
\cortext[cor1]{Corresponding author}
\author[FRIB,MSUPHYS]{G.~Bollen}
\author[FRIB]{K.~Lund}
\author[FRIB,MSUPHYS]{C.~Nicoloff}
\author[FRIB]{S.~Schwarz}
\author[FRIB]{C.~S.~Sumithrarachchi}
\author[FRIB]{A.~C.~C.~Villari}

\address[FRIB]{Facility for Rare Isotope Beams, East Lansing, MI, USA}
\address[MSUPHYS]{Department of Physics and Astronomy, Michigan State University, East Lansing, MI, USA}

\begin{abstract}
Linear gas stoppers are widely used to convert high-energy, rare-isotope beams and reaction 
products into low-energy beams with small transverse emittance and energy spread.  Stopping of the 
high-energy ions is achieved through interaction with a buffer gas, typically helium, generating 
large quantities of He$^+$/e$^-$ pairs.  The Advanced Cryogenic Gas Stopper 
(ACGS) was designed for fast, efficient stopping and extraction of high-intensity, rare-isotope 
beams.  As part of the design process, a comprehensive particle-in-cell code was developed to 
optimize the transport and extraction of rare isotopes from the ACGS in the presence of space 
charge, including He$^+$/e$^-$ dynamics, buffer gas interactions including gas flow, RF carpets, 
and ion extraction through a nozzle or orifice.  Details of the simulations are presented together 
with comparison to experiment when available.

\end{abstract}

\begin{graphicalabstract}
\begin{figure}[h]
\center{
\resizebox{1\textwidth}{!}{%
\includegraphics{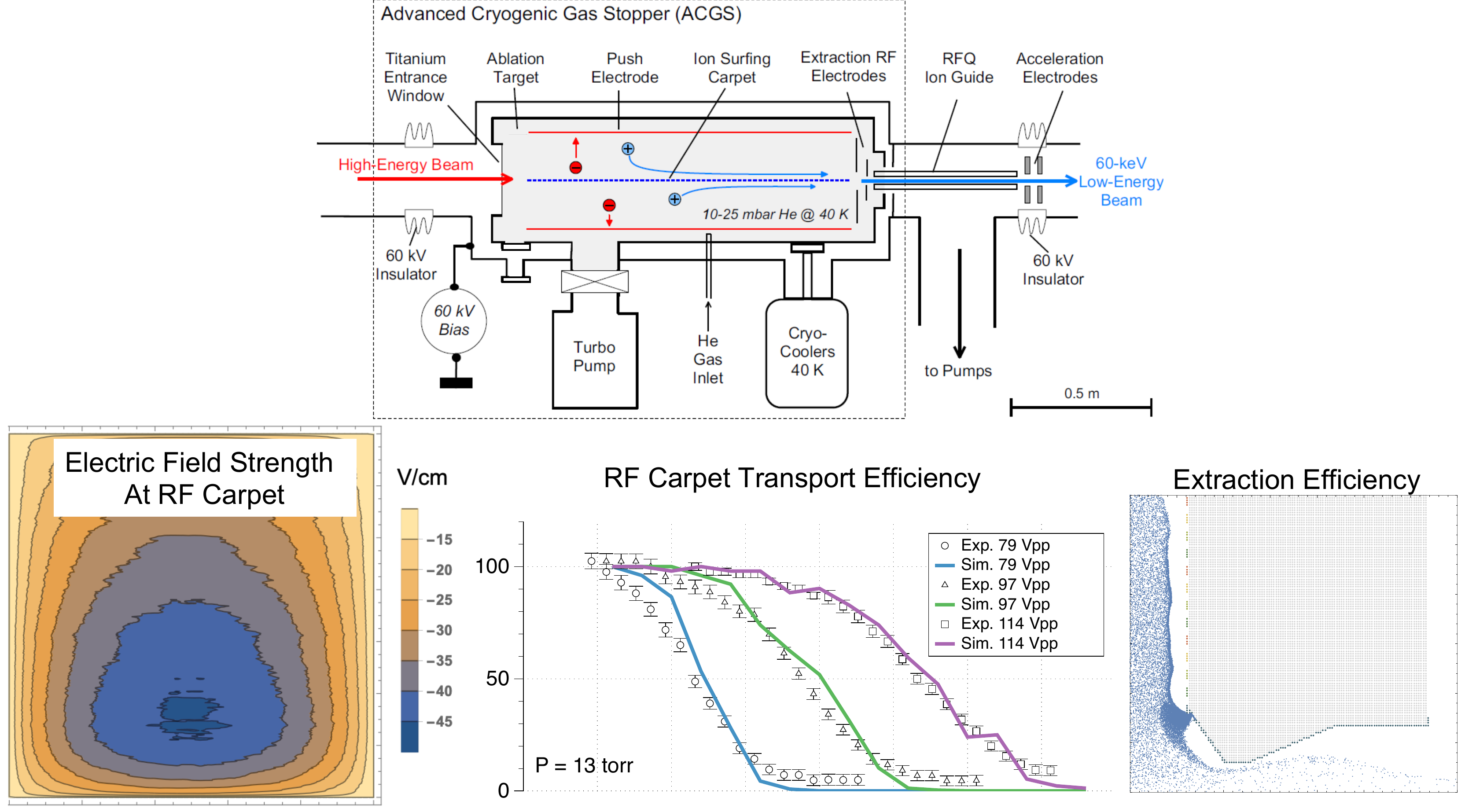}}
}%
\end{figure}
\end{graphicalabstract}

\begin{highlights}
\item 2D and 3D fast particle-in-cell techniques were developed to study the impact of space charge in a gas cell for stopping 
energetic rare isotopes.
\item Codes were compared to experimental results of ion transport across novel wire RF carpets and extraction of ions 
through an orifice with gas flow.
\item Sources of efficiency loss in ion transport up to 10$^8$ incident ions were studied, and bottlenecks in operation 
at higher incident ion intensities were identified.
\end{highlights}


\end{frontmatter}


\section{Introduction}
\label{intro}
Linear gas stoppers have proven instrumental in coupling rare-isotope 
production methods at high energies, up to 100{'}s of MeV/nucleon, to 
high-precision experiments that typically require beam energies on the 
order of 10{'}s of keV/nucleon~\cite{wada2003, Savard2016, Droese:2014ie, 
Petrick:NuclearInstrumentsMethodsInPhysicsResearch:2008, Cooper:2014ji, lund2019}.  
This is typically accomplished in two stages.  
In the first stage the majority of kinetic energy is removed with the 
use of solid degraders.  In the second stage, the rare isotopes are 
essentially brought to rest by subsequently injecting them into a 
chamber filled with a buffer gas.  Helium is typically used since it 
has the highest first ionization energy of any element, 
which minimizes neutralization of the rare isotope ions.  Some gas stoppers 
rely on neutralizing the rare-isotope ions using different buffer gas species, 
followed by re-ionization once they have been extracted, and are not considered 
here.  Once the rare-isotope ions have been stopped, they are guided through the 
chamber using some combination of radio-frequency (RF) and/or static electric fields to an 
extraction region.  The extraction is accomplished through a small 
nozzle or orifice where gas flow provides the transport from the high-pressure 
chamber to the next, lower-pressure, stage of beam delivery.

The particle-in-cell (PIC) technique~\cite{Birdsall:PlasmaPhysicsViaComputerSimulation:1985,
Hockney:ComputerSimulationUsingParticles:1988} has been employed for decades in the 
study of plasmas, gravitational systems, geodynamics, etc.  In the application presented here, 
Poisson's equation is solved numerically on a computational grid, and the electric field at each 
grid point is obtained.  The positions and velocities of simulation particles representing ions 
and electrons are then updated based on the electric field at their positions.  The new particle locations 
are used to update the charge density on the grid at which point the electric field is 
recalculated.  The simulation continues in this fashion until complete.  The 3DCylPIC 
code~\cite{Ringle:2011uf} is a particle-in-cell code that was originally written 
to simulate ion trap and ion transport devices best modeled using a 3D cylindrical coordinate 
system.  It has since been updated to support other 2D and 3D coordinate systems, and was 
used extensively to guide the design of the Advanced Cryogenic Gas Stopper 
(ACGS)~\cite{lund2019}.

\section{The Advanced Cryogenic Gas Stopper}
\label{acgs}

\begin{figure*}[ht]
\center{
\resizebox{0.8\textwidth}{!}{%
\includegraphics{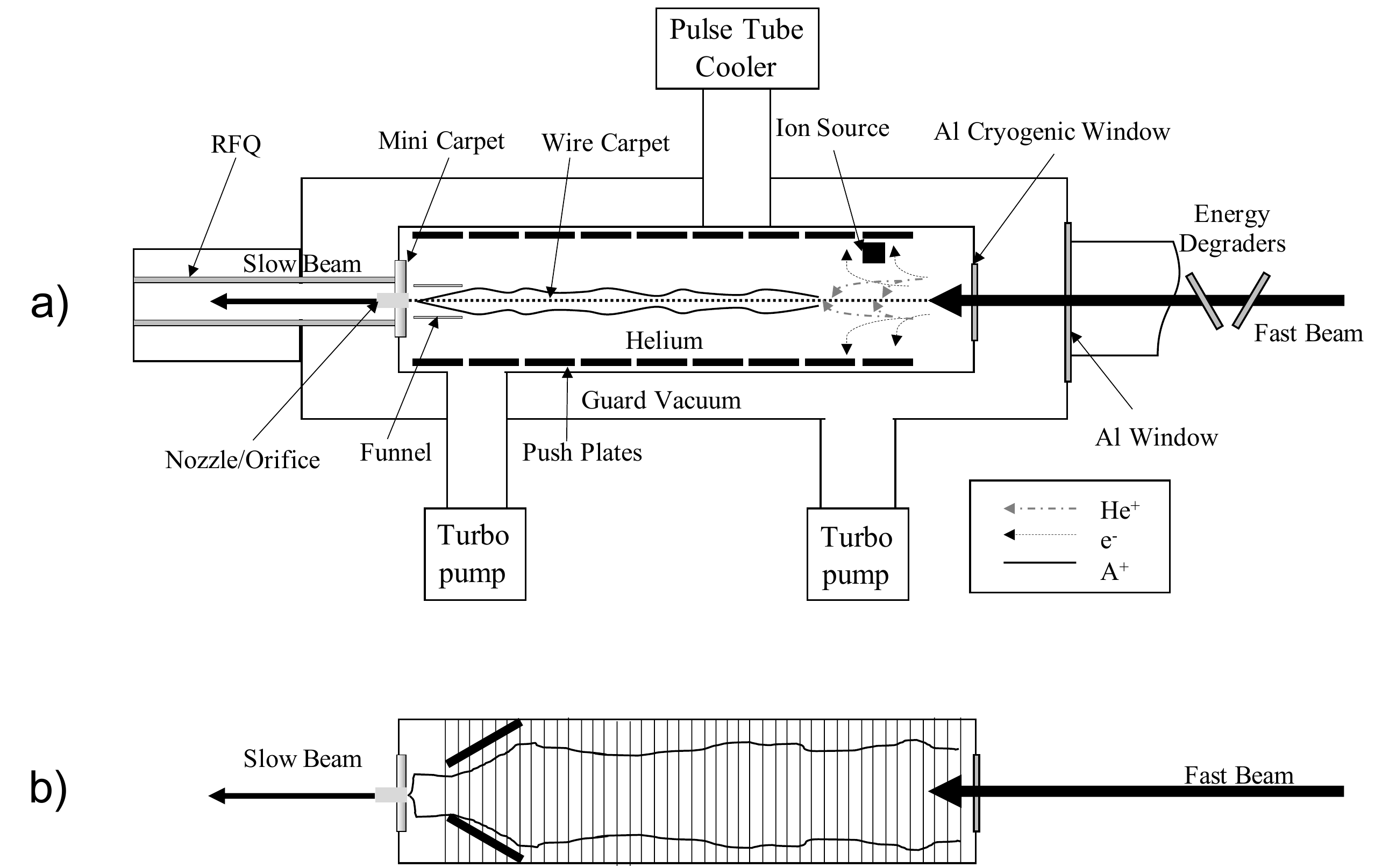}}
}%
\caption{Schematic drawing of the Advanced Cryogenic Gas Stopper.  The major components are 
shown in the side view (a).  The top view (b) illustrates the orientation of the funnel which is 
required to focus the beam onto the mini RF carpet.}
\label{acgs_schematic}
\end{figure*}

\subsection{Concept}
\label{acgs_concept_sec}
The ACGS was conceived~\cite{Bollen2011131} to improve the ionization-dependent efficiency effect that has been 
widely observed~\cite{Cooper:2014ji,Weissman:NuclearInstrumentsMethodsInPhysicsResearch:2005,facina2004,
Takamine:ReviewOfScientificInstruments:2005,neumayr2006} in existing linear gas stoppers, which severely 
limits their performance in delivering high-intensity beams.  The two features that were thought most 
likely to yield the largest positive impact on high-intensity beam delivery were cryogenic operation 
and fast removal of He$^+$ ions.  Cryogenic operation freezes out many possible contaminants in 
the buffer gas, thus reducing molecular ion formation with rare isotopes.  It is also predicted to reduce the number 
of stable molecular ions that will be transported to the end of the ACGS, where they can potentially reduce the 
efficiency of the extraction system.  Fast removal of the He$^+$ ions also serves a dual purpose.  The 
faster the He$^+$ ions are removed, the less likely they are to undergo a charge-exchange reaction with 
contaminants in the buffer gas, and the overall number of He$^+$ ions in the stopping volume is reduced which 
yields a smaller space-charge-generated electric field strength at the surface of the wire RF carpet.

\subsection{Design}
\label{acgs_design_sec}
A schematic of the ACGS is shown in Fig.~\ref{acgs_schematic}.  Fast ions enter from the right with an 
energy of about 1-2 MeV/nucleon prior to the windows.  They pass through two aluminum windows into the cryogenic 
stopping chamber filled with a helium buffer gas.  A central RF carpet bisects the incoming beam in the horizontal 
plane, minimizing the distance that the He$^+$ ions have to travel before impinging and neutralizing on 
the RF carpet.  The RF carpet is made of bare metal wires, 50 \textmu m in diameter with a center-to-center 
distance of 400 \textmu m.  It uses a 4-phase, traveling-wave RF surfing technique~\cite{Bollen2011131,Brodeur2012} 
to transport ions toward the extraction system.  Electrodes, called push plates above and below the central RF carpet create 
an electric field that forces the ions toward the carpet. A potassium ion source is installed in the top rear-most push 
plate for testing and optimizing ion transport and extraction without online beam.

\begin{figure}[h]
\center{
\resizebox{0.5\textwidth}{!}{%
\includegraphics{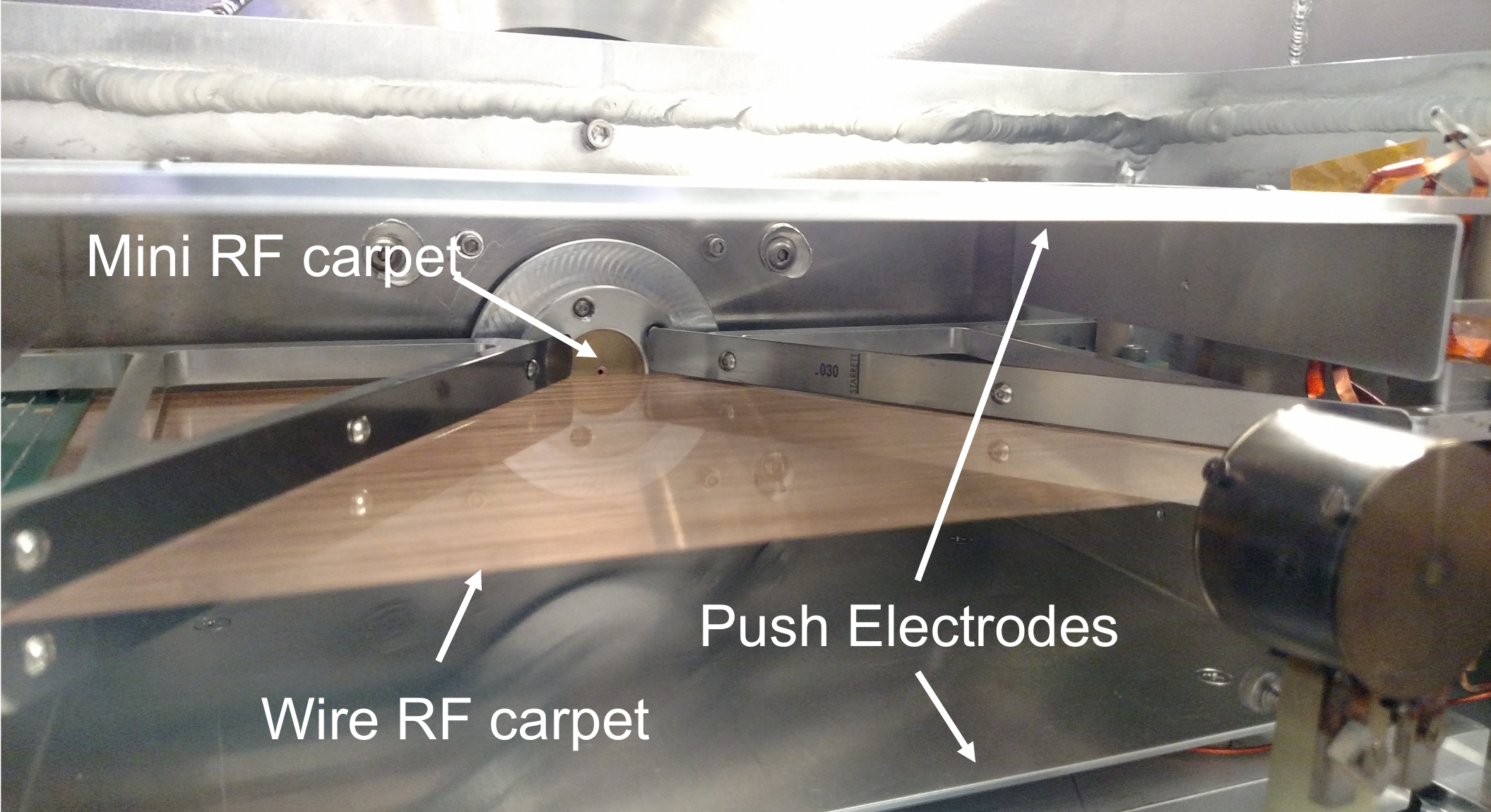}}
}%
\caption{Photo of the final wire RF carpet section, funnel, and mini RF carpet.}
\label{extraction_sys_photo}
\end{figure}

The extraction system of the ACGS consists of a DC funnel and mini RF carpet, and is shown in Fig.~\ref{extraction_sys_photo}.  
The DC funnel is a set of four electrodes that sit just above and below a section of the 
final wire RF carpet section and is used to focus the transported ions onto the mini RF carpet.  The mini 
RF carpet system is composed of a shield electrode, the mini RF carpet structure, and an orifice for extracting the ions 
out of the ACGS and into an RFQ for differential pumping and further transport downstream to experiments.  The mini RF carpet 
is a circular printed circuit board with traces 150~$\mu$m wide separated by a gap of 150~$\mu$m, and runs with at an RF 
frequency of 6.6 MHz with an amplitude of up to 200 V$_{pp}$.  Just like the wire RF carpet sections, the mini RF carpet uses 
a 4-phase traveling wave that transports the ions toward the extraction orifice.  Two extraction orifices with different 
geometries were studied and tested experimentally, based on improved extraction efficiency predictions from 
PIC simulations.  These results will be discussed in Sec.~\ref{extract_pic}.

\subsection{Space Charge and Efficiency}
\label{acgs_spacecharge_sec}
The primary mechanism responsible for the removal of kinetic energy from the fast rare 
isotopes in the helium buffer gas is the creation of He$^+$/e$^-$ pairs at an 
energy cost of 43.5 eV per pair~\cite{Harris:1976tj}.  This can yield millions of He$^+$/e$^-$ 
pairs per incident rare isotope and lead to two effects, as pointed out in \cite{savard2011}, 
that can negatively impact the efficiency of rare isotope extraction from a 
linear gas cell.  The first effect is due to the field created by a large amount of 
positive charge in the gas cell chamber.  Since the mobility of 
electrons~\cite{PhysRev.154.138,ramanan1990electron} is about 100 times higher than 
that of helium ions~\cite{viehland2016} in a helium buffer gas, electrons are quickly 
removed leaving the slower helium ions to build up a large amount of space charge.  
If the removal rate of the helium ions is fast enough, the system will reach 
equilibrium without the formation of a neutral plasma region.  If not, 
then a region of neutral plasma will form that will screen its internal volume from all 
applied RF and static electric fields.  The impact of this effect is discussed in Sec.~\ref{body_pic}.

The second effect relates to buffer gas purity and the generation of molecular ion beams.  
As a result of the large numbers of helium ions created in the stopping process, and its high 
first-ionization energy, the helium ions will preferentially undergo a charge-exchange 
reaction with any trace molecular or atomic contaminant in the buffer gas.  These potentially 
intense contaminant ion beams can then be transported to the extraction region of the gas 
stopper where, due to the high charge density that they generate, they can reduce efficiency 
of the extraction system.  This effect is the topic of Sec.~\ref{extract_pic}.

The total efficiency of the system can be expressed by Eq.~\ref{eff_eq}, where $\epsilon_{stopping}$ is the 
efficiency of rare-isotope ions stopped in the gas relative to the number of rare-isotope ions just prior to 
the energy degraders, $\epsilon_{transport}$ is the efficiency of transporting the ions across the RF carpets, 
and $\epsilon_{extraction}$ is the efficiency of transporting and extracting the ions from the mini carpet.  It 
will be shown that $\epsilon_{transport}$ and $\epsilon_{extraction}$ are negatively impacted by primary and 
secondary effects due to space charge.  Even though $\epsilon_{stopping}$ is not directly impacted by space charge, 
seeking to maximize it by increasing the buffer gas pressure can exacerbate space charge effects that subsequently 
reduce $\epsilon_{transport}$ and $\epsilon_{extraction}$.  Any effects resulting in the neutralization of the 
rare-isotope ions are not considered here.

\begin{equation}
    \label{eff_eq}
    \epsilon_{total} = \epsilon_{stopping}\cdot\epsilon_{transport}\cdot\epsilon_{extraction}
\end{equation}

\section{PIC Simulations of the ACGS}
\label{acgs_pic}
\subsection{Overview}
\label{sim_overview}
\begin{figure*}[ht]
\center{
\resizebox{1\textwidth}{!}{%
\includegraphics{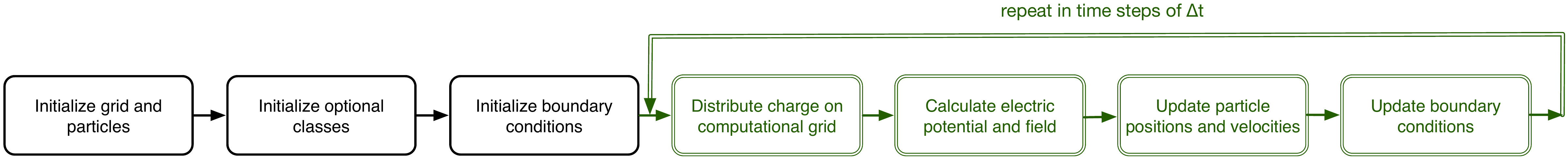}}
}%
\caption{Flow diagram of the PIC code used in the simulations presented here.  Steps in the single-stroke boxes are 
executed once on startup, and the steps in the double-stroke boxes are looped over in time until the simulation is completed.}
\label{pic_struct}
\end{figure*}

An overview of the code used here has already been presented in~\cite{Ringle:2011uf}, so only a brief 
summary and relevant new information will be given.  A diagram illustrating the flow of the PIC 
code is shown in Fig.~\ref{pic_struct}.  First, the grid upon which the electric field is calculated is 
defined.  This includes choosing the geometry, e.g., 2D cylindrical with azimuthal symmetry, 3D Cartesian, 
etc., defining the physical size of the model space covered by grid, and setting the resolution of the 
grid points.  The particle storage classes are also initialized at this point, and provide a structure 
to store information such as position, velocity, mass, charge, etc., for each particle in the simulation.  
Next, any optional classes are initialized.  These typically include routines to handle interactions with 
a buffer gas, or to handle boundary conditions inside of the model space using the capacity matrix 
method~\cite{Hockney:ComputerSimulationUsingParticles:1988}.  Next, the boundary conditions are set to 
their initial values, which are usually associated with potentials applied to electrodes.  

At this point the code enters a loop that is repeated in time steps of \textDelta t, which is determined 
by the time scale of the physics involved, e.g., RF period, particle velocities, mean free path, etc.  
The general practice is to choose the largest value of \textDelta t that doesn't change the results of 
the simulation, and in all simulations described here a \textDelta t~=~5~ns was used.  First, the charge 
associated with the particles in the simulation is distributed using the cloud-in-cell 
technique~\cite{birdsall1969clouds}.  Particles can be created or lost at every time step, so the total 
amount of charge distributed on the grid can vary as the simulation progresses in time.  Next, the electric 
potential is calculated on the grid.  The simulations presented here use spectral methods, such as the one 
described in~\cite{Lai:ImaJournalOfNumericalAnalysis:2002}, as they are very fast and accurate.  After the 
potential has been calculated, the electric field at each grid point is determined by numerical 
differentiation.  Then, the particle positions and velocities are updated using the leap-frog 
method~\cite{Hockney:ComputerSimulationUsingParticles:1988}.  Finally, the time is incremented, boundary 
conditions are updated if necessary, and the loop repeats, continuing until an end condition is satisfied 
or the simulation is manually terminated.

\subsection{Space charge due to creation of He$^+$/e$^-$ pairs}
\label{body_pic}
These simulations use a 3D Cartesian model space with a length of 140~cm (z dimension), a height 
of 5~cm (y dimension), and a width of 10~cm (x dimension), which corresponds to the upper half of 
the ACGS's helium-filled chamber. The number of grid points in the x, y, and z dimensions are N$_x$~=~64, 
N$_y$~=~32, and N$_z$~=~512, producing a single volume element of $\mathrm{dV~\approx~7~mm^3}$.
The boundary conditions are set to produce a uniform electric field of variable magnitude to 
push the He$^+$ ions onto the lower boundary (the location of the RF carpet) and pull the electrons 
onto the upper boundary (the location of the push plate electrodes).  The RF carpet electrode structure is not 
included in this simulation, and it is assumed that any particles making contact with any boundary 
are lost.

At every time step of the simulation a number of He$^+$/e$^-$ pairs must be created that accurately 
represents the energy that is being deposited in the helium buffer gas.  The LISE++~\cite{tarasov2016} 
software package was used to calculate the energy loss and stopped ion distributions of $^{40}$Cl and 
$^{80}$Ge beams in the helium buffer gas with pressures ranging from 10 to 30 torr at a temperature of 
T~=~50~K.  These isotopes were chosen to obtain representative momentum distributions from LISE++ calculations, 
and also due to their A/Q values for singly charged ions, where $^{40}$Cl$^+$ is at the low end of the ACGS design spec, 
and $^{80}$Ge$^+$ is more representative of the majority of stopped beams delivered at the NSCL.  For each 
pressure the energy degraders, shown in Fig.~\ref{acgs_schematic}, were adjusted to obtain the maximum stopping 
efficiency of ions in the buffer gas. 

\begin{equation}
    \label{skewnorm_eq}
    f(a,\mu,\sigma,\alpha)=a\frac{e^\frac{-(z-\mu)^2}{2\sigma^2}\cdot \mathrm{Erfc}\left[\frac{-\alpha(z-\mu)}{\sqrt{2}\sigma}\right]}{\sqrt{2\pi}\sigma}
\end{equation}

Functions were fit to the energy loss and stopped ion data in the transverse (x, y) and axial dimensions (z).  
Probability distributions based on energy loss fits were used to generate random starting positions of the 
He$^+$/e$^-$ pairs along a straight line to represent the ionization trail of a single rare isotope entering 
the ACGS.  Gaussian functions were used to fit the data in the transverse dimensions. A skew normal function, 
of the form given in Eq.~\ref{skewnorm_eq}, was used to fit the data in the axial dimension as it was found to 
describe the data well and is easy to generate random numbers from.  Due to the exceptionally large numbers of 
He$^+$/e$^-$ pairs generated in the stopping process, a charge scaling, Q$_{scale}$ was introduced in order to 
reduce the computational expense of tracking and moving the simulation particles.  This charge scaling 
simply scales, particle by particle, the amount of charge assigned to the grid for calculation of the electric 
field, but leaves the physics of moving the particles unchanged.

\begin{equation}
    \label{scale_eq}
    Q_{scale}=\frac{\bar{E}_{RI}}{E_{pair}\cdot N_{pair}}
\end{equation}

The charge scaling used is given by Eq.~\ref{scale_eq}, where $\bar{E}_{RI}$ is the average energy deposited 
in the helium buffer gas per rare isotope, $\mathrm{E_{pair}}$ is the energy required to create a $\mathrm{He^+/e^-}$ 
pair (43.5 eV~\cite{Harris:1976tj}), and $\mathrm{N_{pair}}$ is the number of random $\mathrm{He^+/e^-}$ pairs generated for 
each rare-isotope-ionization trail.  Each ionization trail was generated with $\mathrm{N_{pair}=200}$, which corresponds 
to $\mathrm{Q_{scale}\approx4644}$.  A random ionization trail was generated every 2000, 200, or 20 ns to achieve an incoming 
rare-isotope rate of $\mathrm{10^6}$, $\mathrm{10^7}$, or $\mathrm{10^8}$ pps, respectively.

Buffer gas interactions in these simulations were modeled using the SDS 
method~\cite{Appelhans:InternationalJournalOfMassSpectrometry:2005}, which combines viscous mobility 
with random jumping of particles to reduce the computational overhead compared to using a  collision-by-collision 
method~\cite{Appelhans:InternationalJournalOfMassSpectrometry:2002}.  The force used in conjunction with the leap-frog 
method to advance the particle velocities and positions is given by Eq.~\ref{force_eq} as

\begin{equation}
    \label{force_eq}
    \vec{F}=q\cdot\vec{E}-m\cdot\delta\cdot\vec{v}
\end{equation}
where $q$ is the charge, $\vec{E}$ is the electric field at the particle's location, $m$ is the 
particle's mass, $\delta$ is the damping coefficient, and $\vec{v}$ is the particle's velocity.  The 
damping coefficient is defined in Eq.~\ref{damping_eq} as

\begin{equation}
    \label{damping_eq}
    \delta=\frac{q}{m}\frac{1}{k_\circ}\frac{P/P_n}{T/T_n}
\end{equation}
where $k_\circ$ is the particle's reduced mobility in a particular gas, $P_n$~=~760~torr is the 
normal pressure, and $T_n$~=~293.15~K is the normal temperature.  The reduced mobilities used in these 
simulations are $\mathrm{k_\circ}$~=~1000~$\mathrm{cm^2/(V\cdot s)}$ for electrons~\cite{schwarz1978} and 
$\mathrm{k_\circ}$~=~10~$\mathrm{cm^2/(V\cdot s)}$ for He$^+$ ions in a helium buffer 
gas~\cite{viehland2016}.

\begin{figure}[h]
\center{
\resizebox{0.48\textwidth}{!}{%
\includegraphics{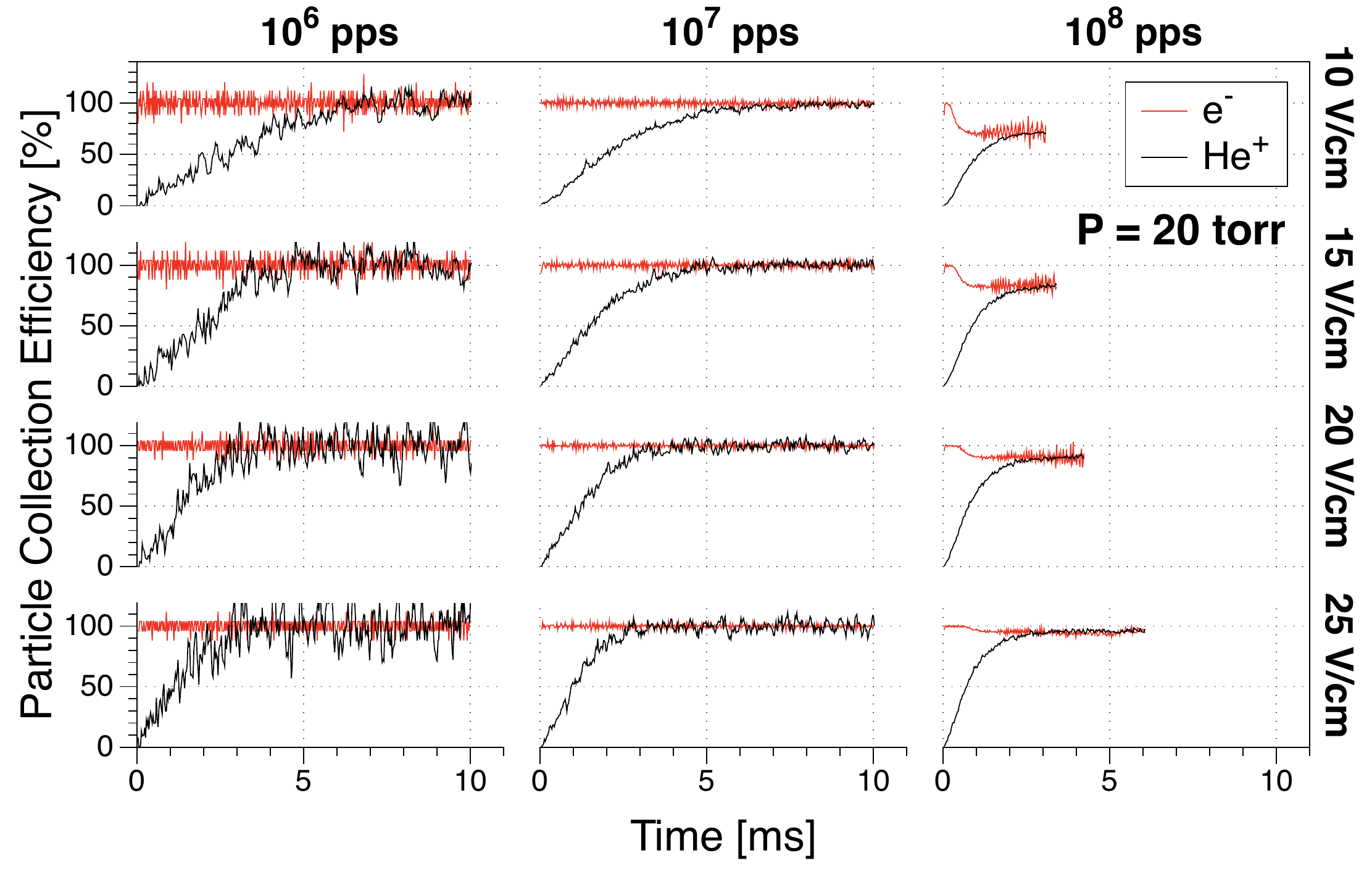}}
}%
\caption{Helium ion (black lines) and electron (red lines) collection efficiencies vs. time for 
different incoming rates of $^{40}$Cl with a helium buffer gas pressure P = 20 torr at T~=~50~K, 
and applied push fields of E$_p$ = 10, 15, 20, and 25 V/cm.}
\label{coll_eff}
\end{figure}

Fig.~\ref{coll_eff} shows the collection efficiency of the He$^{+}$ ions and electrons on the boundaries 
of the model space vs. time for multiple incoming rates of $^{40}$Cl.  In this example, the helium buffer 
gas pressure was P~=~20~torr, and the magnitude of the applied push field was E$_p$~=~10 to 25~V/cm.  
The electron collection efficiency quickly reaches 100\% due to its high mobility, whereas the He$^+$ ion 
collection efficiency grows more slowly.  In all cases, except for those with an incoming particle rate of 
10$^8$~pps in the right-most column, both the He$^+$ ion and electron 
collection efficiencies reach 100\% and a steady state is achieved.  For the cases shown in the right column, 
the electron collection efficiency grows to 100\% for a short time and then decreases.  
The He$^+$ ion collection efficiency also reaches the same collection efficiency as the electrons, and thus a 
steady state is achieved.  The difference in these cases is that a neutral plasma has formed in the stopping volume 
when the positive charge density from the He$^+$ ions is sufficient to begin trapping electrons that 
are generated in that region.  One can see that, by increasing E$_p$, the particle collection efficiency 
increases closer to 100\%, but a value larger than 25~V/cm is required in order to stop the formation of the 
neutral plasma in the case of an incoming particle rate of 10$^8$ pps.

\begin{figure}[h]
\center{
\resizebox{0.48\textwidth}{!}{%
\includegraphics{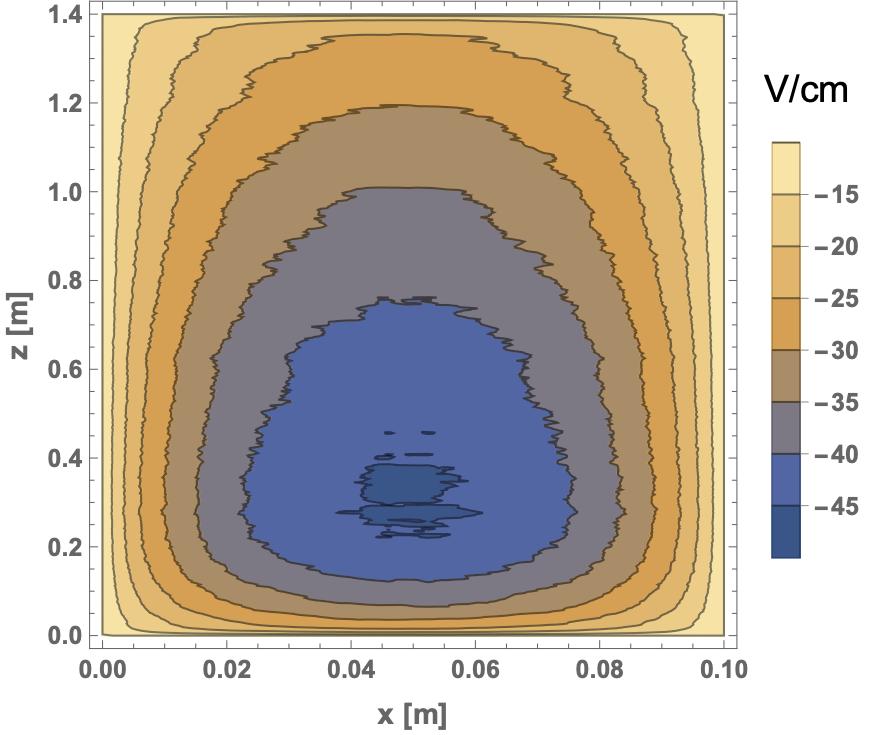}}
}%
\caption{Contour plot of the of the y-component of the electric field at the surface of the RF carpet for 
an incident rate of 10$^{8}$ pps of $^{40}$Cl with an applied push field strength of 10 V/cm and a pressure 
of P = 20 torr at T = 50 K.}
\label{pot_at_carpet}
\end{figure}

Once a steady state of particle collection has been reached, the total potential, due to the applied push 
field and additional space charge potential, stops evolving.  At this point it is instructive to inspect 
the electric field strength at the surface of the RF carpet as this will have a significant impact on 
$\epsilon_{transport}$. As an example, Fig.~\ref{pot_at_carpet} shows a contour plot of the y-component 
of the electric field strength at the surface of the RF carpet for the case of an incident 
beam rate of 10$^{8}$ pps of $^{40}$Cl with a pressure of P~=~20~torr at a temperature of T~=~50~K.  Although 
the magnitude of the applied push field is only E$_{p}$~=~10~V/cm, the magnitude of the total electric 
field strength exceeds 45~V/cm in the area of greatest energy deposition by the incident $^{40}$Cl beam.  
Electric field strength maps will be used in the next section to estimate $\epsilon_{transport}$ 
under a variety of conditions for different isotopes.

If a plasma has formed, and steady-state collection has been reached, the volume that the plasma occupies 
can be estimated by examining the magnitude of the electric field at each grid point.  Since the electric 
field inside the plasma should be zero, one simply counts the grid points with a field strength below a threshold 
and multiplies it by the volume of single element, dV.  Using the stopped ion distribution information extracted from 
the LISE++ calculations, it is also straightforward to estimate the fraction of the rare isotopes that are stopped 
within the plasma volume, and most likely lost.

\begin{figure}[h]
\center{
\resizebox{0.48\textwidth}{!}{%
\includegraphics{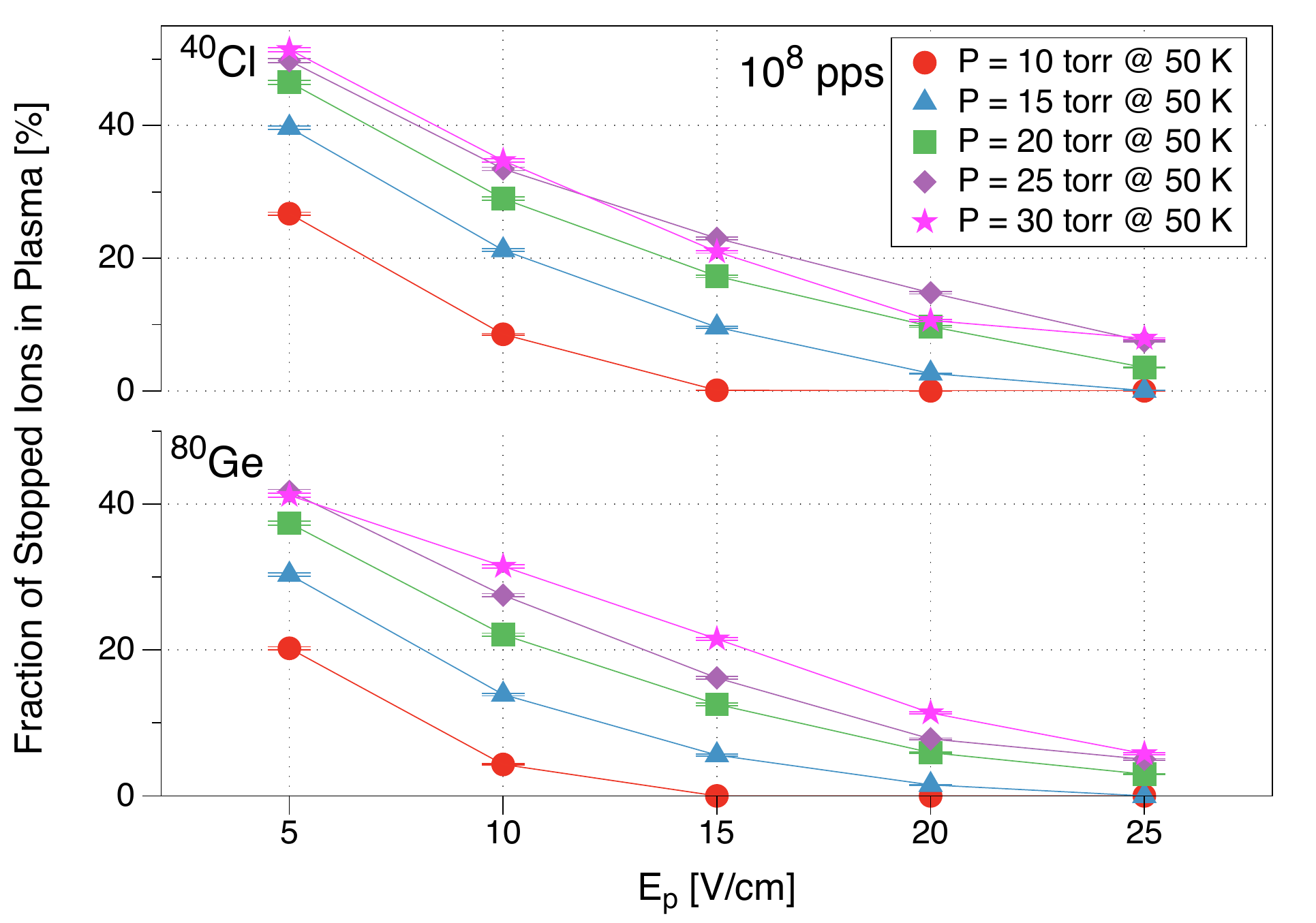}}
}%
\caption{Calculated fraction of stopped rare isotopes within the plasma volume for an incoming rare-isotope rate 
of 10$^8$ pps as a function of E$_p$.  Lines are to guide the eye.}
\label{plasma_ion_fig}
\end{figure}

Fig.~\ref{plasma_ion_fig} shows the overlap of the stopped rare isotopes, $^{40}$Cl and $^{80}$Ge, with the plasma 
volume for different buffer gas pressures and push field strengths for an incoming rate of 10$^8$ pps.  The trends 
are similar for both rare-isotope species in that lower push field strengths and higher buffer gas pressures lead to 
a larger fraction of the rare isotopes stopping within the plasma volume.  This behavior conforms to expectations as 
a higher buffer gas pressure reduces the He$^+$ ion speed and slows their removal.  The same holds true for lower push 
field strengths.  The overlap for $^{40}$Cl is slightly higher than that for $^{80}$Ge, which is probably due 
to a combination of differences in the energy deposition distributions and the slightly higher total energy per particle 
deposited by the $^{40}$Cl ions.

\subsection{Ion transport efficiency of a traveling wave RF carpet}
\label{rf_carpet_trans}
\begin{figure}[h]
\center{
\resizebox{0.48\textwidth}{!}{%
\includegraphics{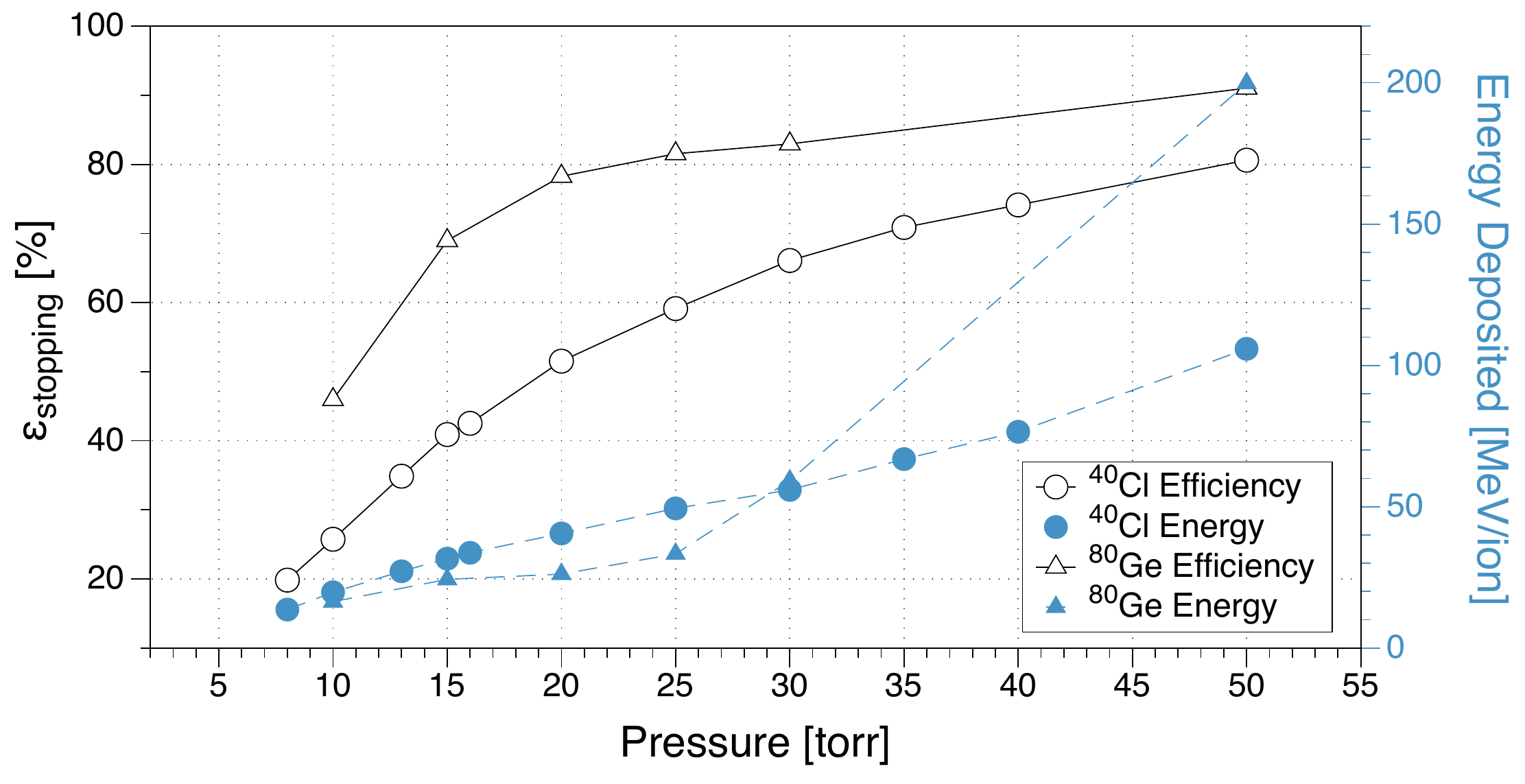}}
}%
\caption{Stopping efficiency, $\epsilon_{stopping}$, and energy deposited per incoming ion as a function 
of buffer gas pressure for $^{40}$Cl and $^{80}$Ge beams.}
\label{LISE_eff}
\end{figure}

The buffer gas pressure, total electric field at the RF carpet surface, and applied RF amplitude and 
frequency are the primary drivers of $\epsilon_{transport}$ of a traveling wave RF carpet.  
Fig.~\ref{LISE_eff} shows the efficiency and energy deposited per incoming ion for two beams.  As 
expected a higher gas pressure yields a higher $\epsilon_{stopping}$.  However, this comes at a 
cost of increasing the amount of energy deposited in the buffer gas.  This is due to the interplay of 
the energy degraders and buffer gas pressure on efficiency.  For thicker energy degraders there are 
more ion losses in the degraders and the two entrance windows.  Increasing the gas pressure permits the reduction of the 
degrader thickness and increases the stopping efficiency.  Simply increasing the buffer 
gas pressure in order to increase the stopping efficiency may not always be the best approach as it 
leads to greater energy deposition (more He$^+$/e$^-$ pairs), and also slows the collection of the 
He$^+$ ions.  Both of these effects result in greater electric fields due to space charge at the surface 
of the RF carpet, which can lower the $\epsilon_{transport}$.

Using the internal potassium surface ion source in the ACGS, measurements of the transport efficiency 
across two sections (about 300 mm) of the central wire RF carpet were performed.  For these measurements, the 
ion current produced from the source was reduced to the point where space-charge effects were negligible.  
The ions were collected on the funnel electrode and a transport efficiency of 100\% was achieved across 
multiple carpet segments in these measurements for applied push field strengths of E$_p$~=~5~V/cm.  This 
was used to calibrate a baseline current after which the potential applied to a push plate electrode was 
increased and the change in current was recorded.  The buffer gas pressure and RF amplitude were also varied 
and the traveling wave was generated using four independent phases, each with an RF frequency of 4 MHz. 
 
\begin{figure}[h]
\center{
\resizebox{0.5\textwidth}{!}{%
\includegraphics{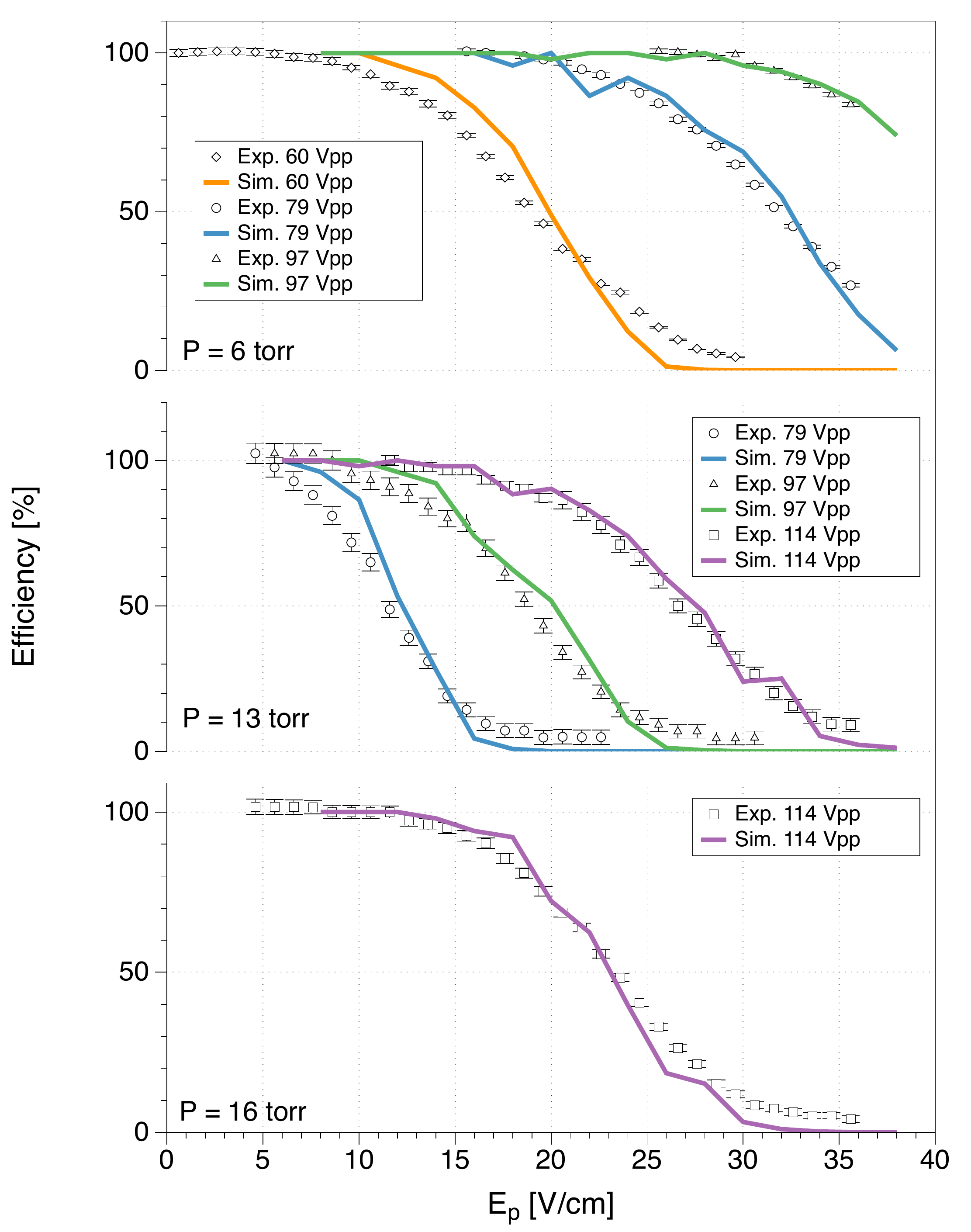}}
}%
\caption{Transport efficiency, $\epsilon_{transport}$, of $^{nat}$K$^+$ ions across two wire 
RF carpet sections (300 mm linear distance) of the ACGS as function of E$_p$ for 
various RF amplitudes and buffer gas pressures.  Simulation results from IonCool package are shown as solid 
lines.}
\label{carpet_eff_meas}
\end{figure}

\begin{figure}[h]
\center{
\resizebox{0.48\textwidth}{!}{%
\includegraphics{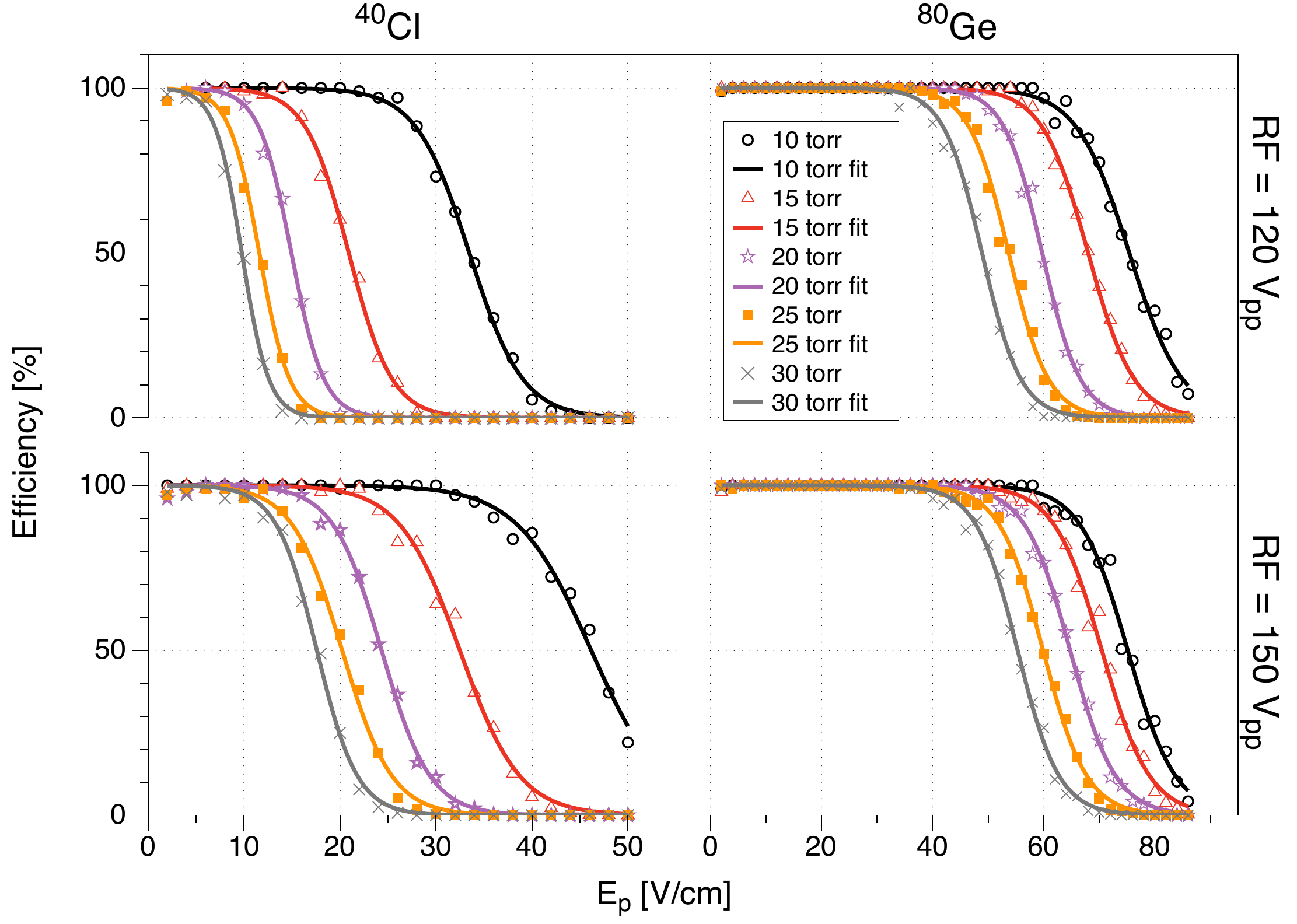}}
}%
\caption{IonCool simulations of transport efficiency as a function of E$_p$ across two segments 
of ACGS wire RF carpets for $^{40}$Cl and $^{80}$Ge with RF amplitudes of 120 V$_{pp}$ and 150 V$_{pp}$.  Solid lines 
are fits of sigmoid functions to the data.}
\label{ioncool_surf_eff}
\end{figure}

The results of the study are given in Fig.~\ref{carpet_eff_meas}.  The three plots show the transport 
efficiency as a function of electric field strength for an equivalent pressure of P = 6, 13, and 16 torr 
at a temperature of T = 50 K for different RF amplitudes. The primary feature to note is that an increase in 
pressure reduces the push field tolerance for a given RF amplitude.  Solid lines are simulation results 
obtained using the IonCool package~\cite{schwarz2006}, and show good agreement with the measurements.  
Since the agreement is good, we can confidently use IonCool to predict the push field tolerance under 
a variety of conditions.  This is shown in Fig.~\ref{ioncool_surf_eff} for our two test-case isotopes of 
$^{40}$Cl and $^{80}$Ge, where the transport efficiency across two ACGS wire RF carpet modules, with a total 
length of 300~mm, is simulated as a function of the push field strength for various buffer gas pressures and 
two RF amplitudes.  The solid lines are fits of sigmoid functions, which will be used in later simulations, 
given by Eq.~\ref{sigmoid_eq}, where $a$ and $b$ are free parameters.  The lower RF amplitude of 120 V$_{pp}$ 
is the current maximum value used in ACGS to avoid igniting a discharge.  A safe value of 150 V$_{pp}$ was 
achieved with an RF carpet assembly in a test stand for the buffer gas pressure range of interest, and it is 
believed that the wiring path of the RF carpets in the ACGS is to blame for the reduced performance, which has 
a significant impact on $\epsilon_{transport}$ for higher push field strengths for lower-mass isotopes.  

\begin{equation}
    \label{sigmoid_eq}
    \epsilon=\frac{100}{1+e^{a(E_p-b)}}
\end{equation}

The reduced push-field tolerance of the $^{40}$Cl$^+$ ions vs. $^{80}$Ge$^+$ ions can be understood as a reduction in 
the strength of the pseudo potential arising from the RF carpet \cite{schwarz2011,Bollen2011131}, and is given by 
\begin{equation}
    \label{pseudopotential_eq}
    V_{eff,d}=\frac{\omega_{RF}^2}{\omega_{RF}^2+\left(q/(mK_o)\right)^2}\frac{q}{4m\omega_{RF}^2}\overline{E^2},
\end{equation}
where $\omega_{RF}$ is the angular frequency of the applied RF, $q$ is the ion charge, $K_o$ is the reduced ion mobility, 
$m$ is the ion mass, and $\overline{E}$ is the electric field strength from the RF carpet.  For the operating conditions 
given here, V$_{eff,d}$($^{80}$Ge$^+$)~$\approx$~2$\cdot$V$_{eff,d}$($^{40}$Cl$^+$), resulting in a higher push-field 
tolerance of the $^{80}$Ge$^+$ ions.

\begin{figure}[h]
\center{
\resizebox{0.48\textwidth}{!}{%
\includegraphics{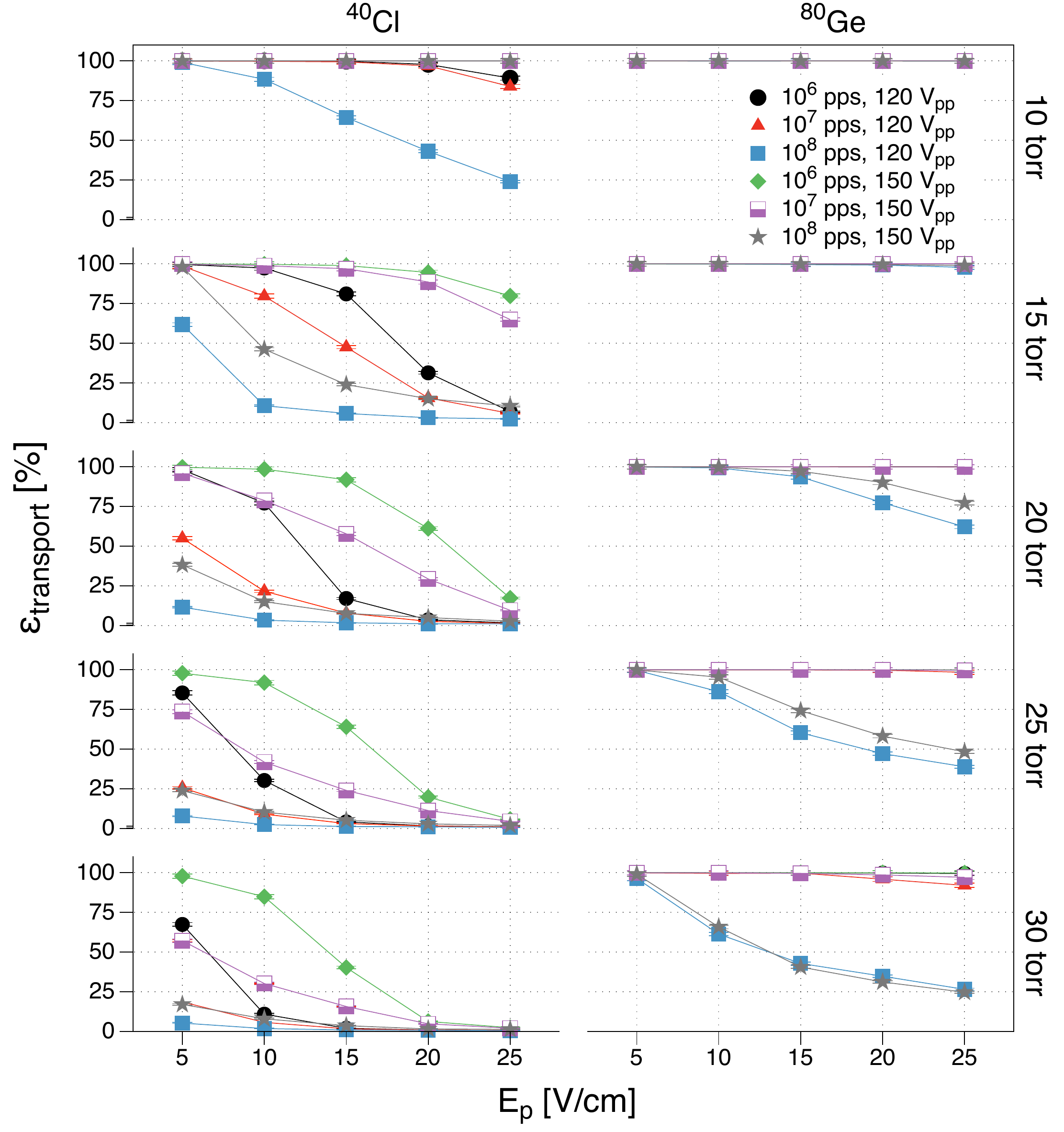}}
}%
\caption{$\epsilon_{transport}$ vs. E$_p$ for $^{40}$Cl$^+$ and $^{80}$Ge$^+$ ions with incoming rates of 
10$^6$ to 10$^8$ pps.  Results for helium buffer gas pressures of P~=~10-50 torr at T~=~50~K and two RF amplitudes, 
V$_{RF}$~=~120~V$_{pp}$ and 150~V$_{pp}$, are shown.  Lines are to guide the eye.}
\label{carpet_trans_eff}
\end{figure}

Combining the push field tolerance simulations (from IonCool), the distribution 
of the stopped rare isotopes (from LISE++), and the total electric field on the carpet surface (from PIC),
gives us a method to estimate $\epsilon_{transport}$.  Using the Monte Carlo technique, random starting positions on 
the RF carpet surface were generated and the probability of survival was calculated from the starting position 
to the end of the model space for 5000 ions.  The results are shown in Fig.~\ref{carpet_trans_eff} for incoming rates 
of 10$^6$ to 10$^8$ pps of $^{40}$Cl and $^{80}$Ge for various buffer gas pressures and RF amplitudes.  As expected, 
the reduced efficiency vs. push field for transport of $^{40}$Cl$^+$ ions, relative to the heavier $^{80}$Ge$^+$, yields 
a lower $\epsilon_{transport}$. Increasing the RF amplitude applied to the RF carpets has a significant impact on 
$\epsilon_{transport}$ for $^{40}$Cl, and is key to optimizing $\epsilon_{transport}$ for light species in general.  This 
behavior can also be exploited if the isotope of interest is significantly heavier than the contaminant ion species.  In 
ACGS the primary stable beam contaminant is O$_2^+$ and its intensity can reach several nanoamps, which can lower 
$\epsilon_{extraction}$ if it is efficiently transported to the end of the system.  The push field strength and RF amplitude 
can be adjusted to lower $\epsilon_{transport}$ for O$_2^+$ while maintaining a high $\epsilon_{transport}$ for a heavy rare 
isotope of interest, increasing the total efficiency for heavier rare isotope beams.

\begin{figure}[h]
\center{
\resizebox{0.48\textwidth}{!}{%
\includegraphics{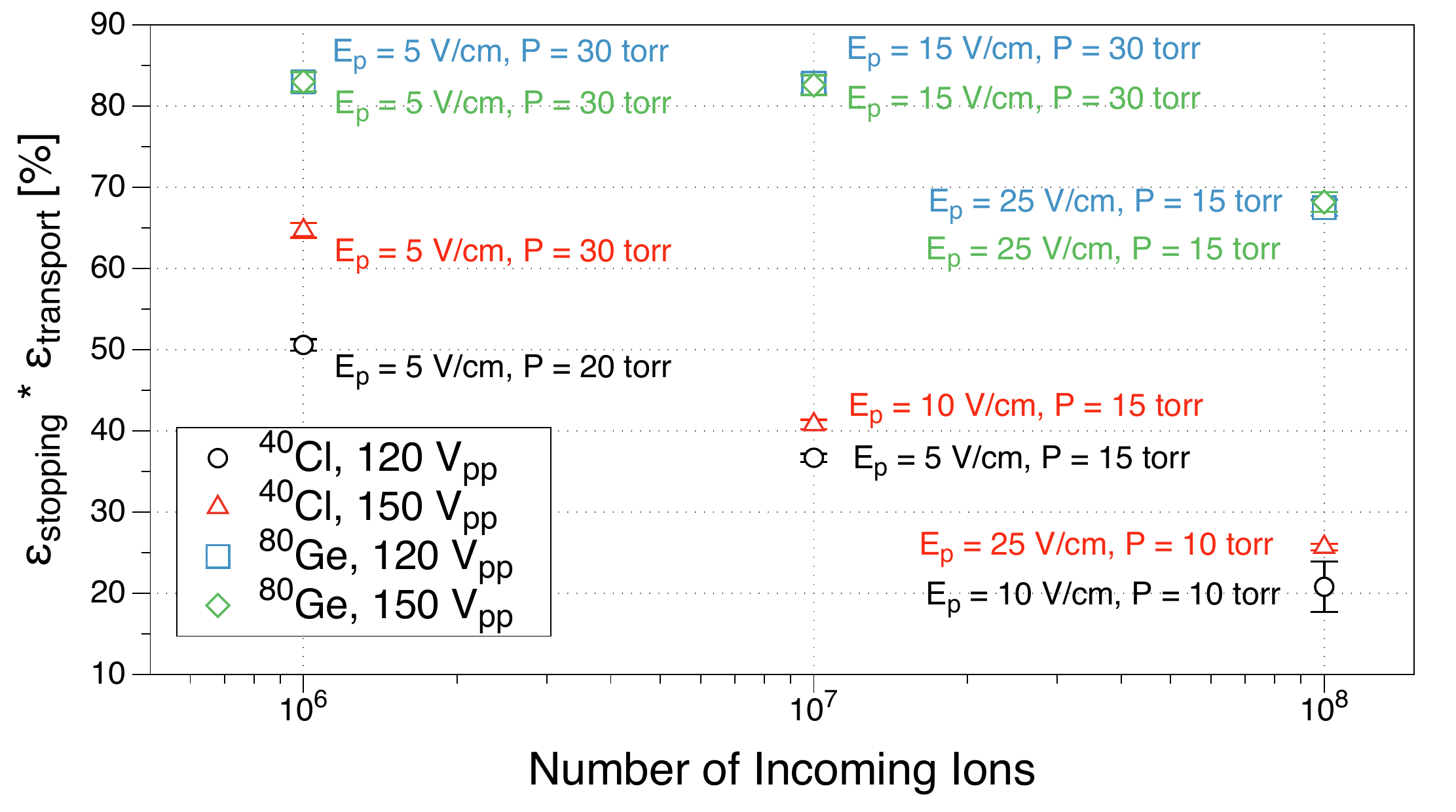}}
}%
\caption{Optimal value of $\epsilon_{transport}$ * $\epsilon_{stopping}$ as a function of the number of incoming ions of 
$^{40}$Cl and $^{80}$Ge for two RF amplitudes, V$_{RF}$~=~120~V$_{pp}$ and 150~V$_{pp}$.  The optimal push field strengths, 
E$_p$ and buffer gas pressures, P, at a temperature of T~=~50~K, are indicated as labels.}
\label{total_trans_eff}
\end{figure}

Using the $\epsilon_{transport}$ data shown in Fig.~\ref{carpet_trans_eff} together with the $\epsilon_{stopping}$ data 
from the LISE++ calculations and the ions lost in any generated plasma from Fig.~\ref{plasma_ion_fig}, the conditions for 
obtaining the optimal $\epsilon_{transport}$ * $\epsilon_{stopping}$ value can be determined, and is shown in 
Fig.~\ref{total_trans_eff}.  The buffer gas pressure and push field strength were chosen such that the maximum efficiency was 
obtained for each point, and are given as labels in the figure.  For the case of $^{40}$Cl the maximum value of 
$\epsilon_{transport}$ * $\epsilon_{stopping}$ is approximately 66\%, and is only realized for an incident rate of 10$^6$ 
pps and a V$_{RF}$~=~150~V$_{pp}$.  In order to maintain the highest efficiency for 10$^7$ and 10$^8$ pps the buffer gas 
pressure must be reduced and the push field strength increased.  Reducing the buffer gas pressure negatively impacts 
$\epsilon_{stopping}$, but it increases the speed of He$^+$/e$^-$ removal which both reduces electric field strength due 
to space charge at the surface of the RF carpet and inhibits the formation of a plasma.  Increasing E$_p$ certainly can have 
a negative impact on $\epsilon_{transport}$, but also inhibits the formation of plasma.  Although the increase in RF 
amplitude improves the total efficiency for 10$^7$ and 10$^8$ pps, the impact is not tremendous and indicates that further 
increases in efficiency would require future RF carpet development and/or reduction in charge density by increasing the 
size of the stopping volume or potentially using multiple layers of RF carpets.  For the case of $^{80}$Ge, the maximum 
efficiency is nearly realized for 10$^6$ and 10$^7$ pps, although moving from 10$^6$ to 10$^7$ pps requires increasing E$_p$ 
from 5 V/cm to 15 V/cm in order to avoid losses associated with ions stopping within the plasma volume.  Increasing the 
incident rate from 10$^7$ to 10$^8$ pps reduces the total efficiency slightly and requires an increase in E$_p$ and a 
reduction in the buffer gas pressure, again to mitigate plasma formation.  Again, in order to maintain high stopping and 
transport efficiencies for rates beyond 10$^7$ for light isotopes (A$\le$40) and rates beyond 10$^8$ for heavier isotopes 
(A$\ge$80) larger stopping volumes with lower buffer gas pressures, faster He$^+$/e$^-$ removal, and advanced 
(e.g., multi-layer) RF carpet technologies are required.  However, this must be complemented with an extraction system 
that can handle potentially significant amounts of ionic stable beam contamination that is generated via charge exchange with 
the He$^+$ ions, and will be explored in Sec.~\ref{extract_pic}. 

\subsection{Impact of space charge on extraction system efficiency}
\label{extract_pic}

\subsubsection{Description of PIC model of extraction system}
\begin{figure}[h]
\center{
\resizebox{0.5\textwidth}{!}{%
\includegraphics{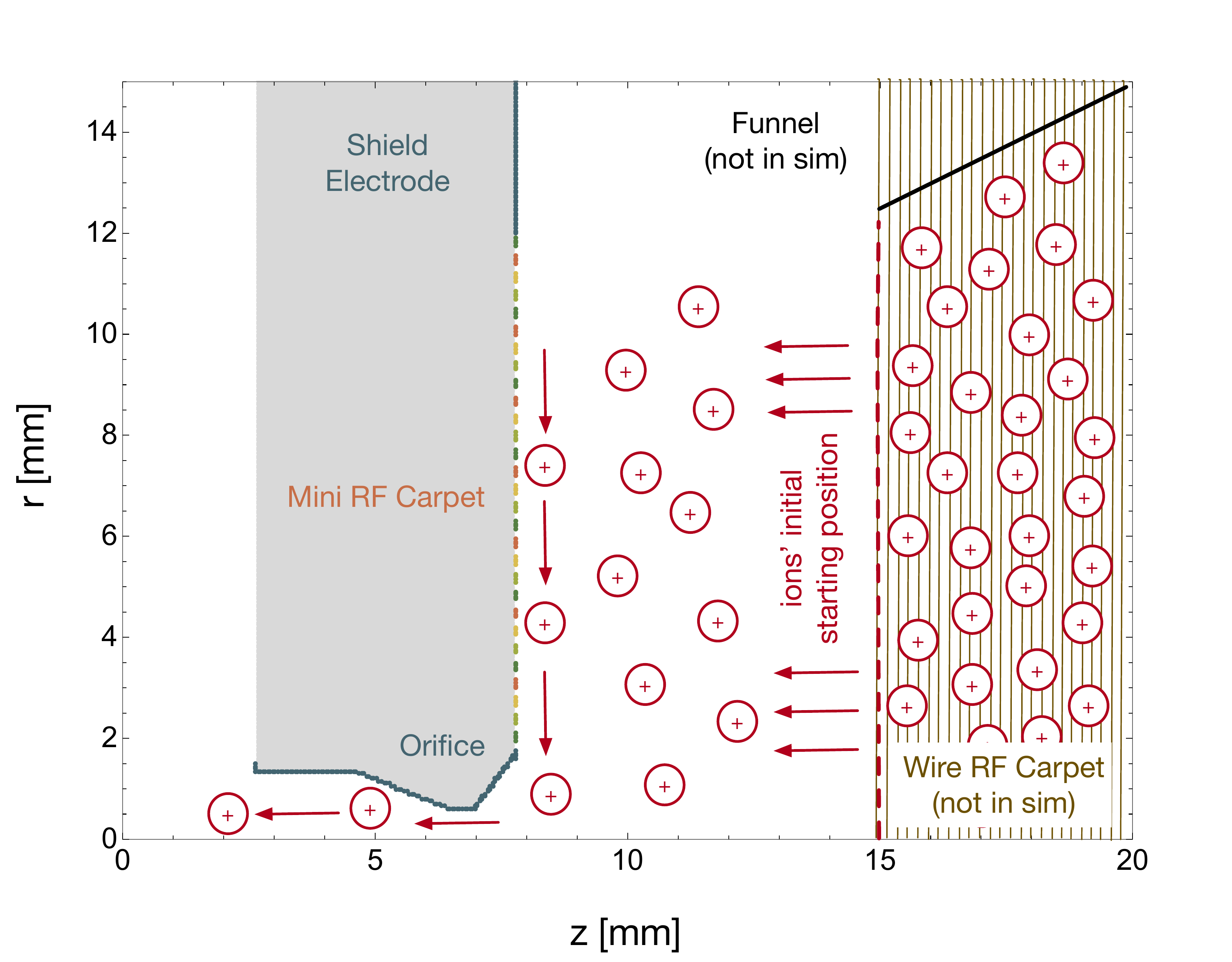}}
}%
\caption{PIC model of the ACGS mini RF carpet with the original extraction orifice.  The DC funnel and wire 
carpet are included for illustration, but not included in the simulation.}
\label{mini_carpet_PIC_model}
\end{figure}

\begin{figure}[h]
\center{
\resizebox{0.5\textwidth}{!}{%
\includegraphics{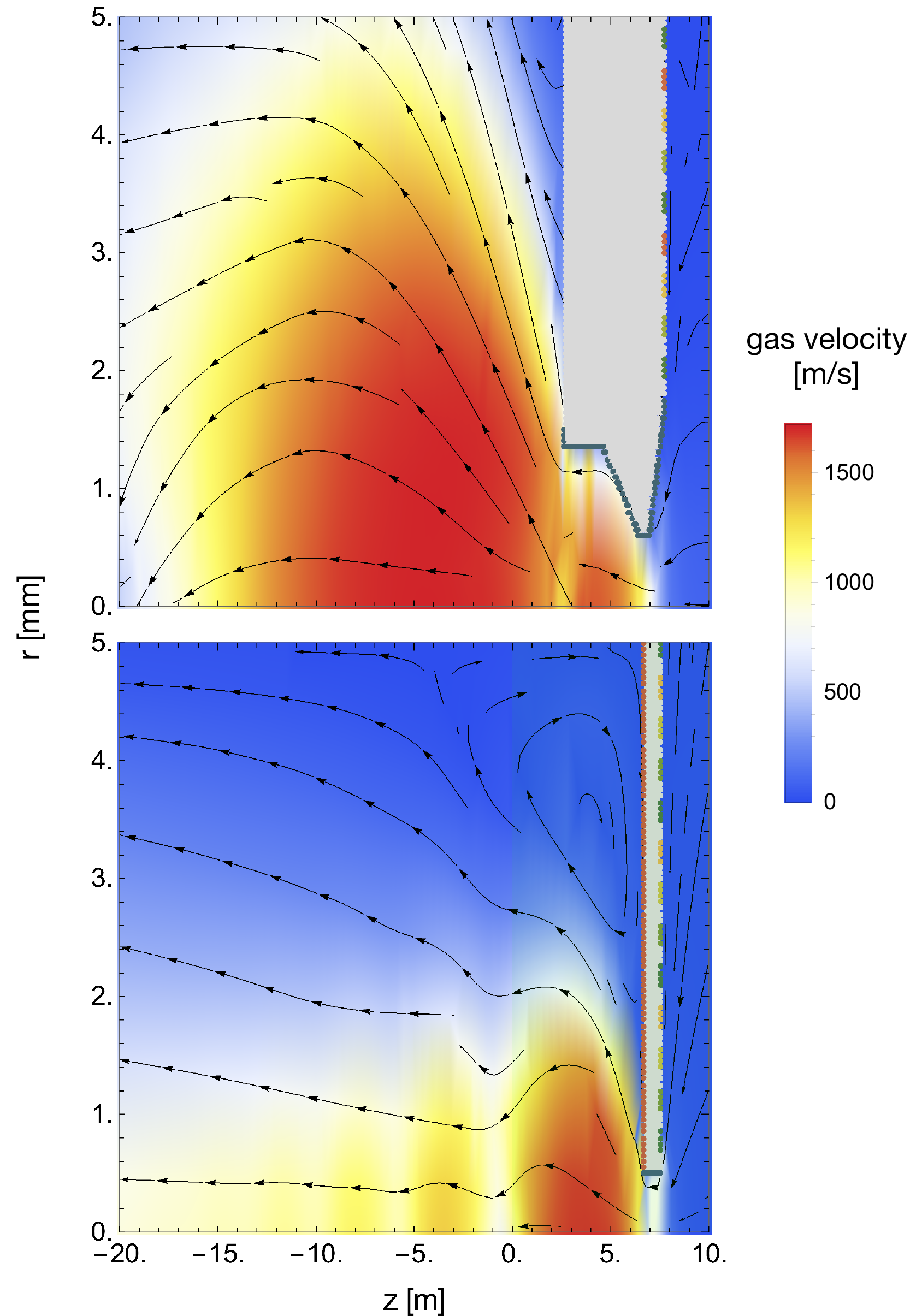}}
}%
\caption{Flow plots of COMSOL calculations of gas flow through the original orifice (top) and simple hole geometry 
that replaced it (bottom) with a buffer gas pressure of P~=~75~torr at temperature of T~=~300~K.}
\label{gas_flow_fig}
\end{figure}

Unlike the PIC simulations discussed previously, the model of the extraction orifice uses a 2D cylindrical 
space with azimuthal symmetry, and is illustrated in Fig.~\ref{mini_carpet_PIC_model}.  Here the wire RF carpet 
and funnel are included for context and are not included in the simulation.  Since electrodes are being 
included in the model space, the capacity matrix method, introduced above, is required.  The 2D model space has a 
radius of 37.2~mm and a length of 20~mm.  The number of grid points in the r and z dimensions are N$_r$~=~745 and 
N$_z$~=~401, respectively, producing annular volume elements with a cross-sectional area of $\mathrm{dA\approx0.0025~mm^2}$.
Ions are created at 
the dashed line at a rate of two ions every timestep. In this case the charge scale used is given by Eq.~\ref{scale_2_eq},
\begin{equation}
    \label{scale_2_eq}
    Q_{scale}=\frac{I\cdot \Delta t}{e\cdot N_{particles}}
\end{equation}
where $I$ is the desired incident ion current, $\Delta t$ is the timestep, $e$ is the elementary charge, and $N_{particles}$ 
is the number of particles created per timestep.  A $\mathrm{Q_{scale}}$ of 15.6 to 312 was to used in these simulations to 
generate incident ion currents ranging from 1 nA to 20 nA.
A constant electric field in negative z direction pushes the ions towards the mini RF carpet where the 
traveling RF wave transports them towards the orifice.  The left-most boundary of the model space corresponds to 
the location of the RFQ, and ions reaching this boundary are considered to be successfully transported.  The 
significant gas pressure differential across the orifice creates a gas flow which transports the ions out of the 
ACGS.  The specifics of the gas flow heavily depend on the pressure and temperature of the gas, and on the geometry 
of the orifice.  COMSOL was used to calculate the steady-state temperature, pressure, and velocity of the gas for 
all cases presented here, and an example is shown in Fig~\ref{gas_flow_fig}.  The results from COMSOL calculations 
were incorporated into the PIC simulations by interpolating the final calculated gas temperature, pressure, and velocity 
onto the PIC grid, and the SDS collision model was used for interactions with the buffer gas, including the bulk flow 
of the gas.

\subsubsection{Comparison of PIC simulation results to experiment}
\label{pic_vs_exp_sec}
During online 
experiments significant stable ion beam currents, up to several nanoamps and primarily O$_{2}^+$, were 
observed.  Assuming that the stable beam is produced via charge exchange with the helium ions created 
during stopping of the rare isotopes, the stable beam current should scale linearly with the incoming rate 
of rare isotopes.  However, during online experiments it was observed that the linear scaling only held 
at low incoming rates and would eventually saturate.  One possibility is that the space charge from the 
stable ions causes a loss in $\epsilon_{extraction}$.  To investigate this, a surface ion source 
inside of the ACGS was used to produce beams of $^{nat}$K$^+$ ions.  The heating current of the ion source was 
adjusted to control the ion beam current, and $\epsilon_{extraction}$ was determined by first measuring the 
incident ion current on the mini RF carpet with no RF, and afterwards on the RFQ with the mini RF carpet tuned 
for optimal transmission. 

\begin{figure}[h]
\center{
\resizebox{0.5\textwidth}{!}{%
\includegraphics{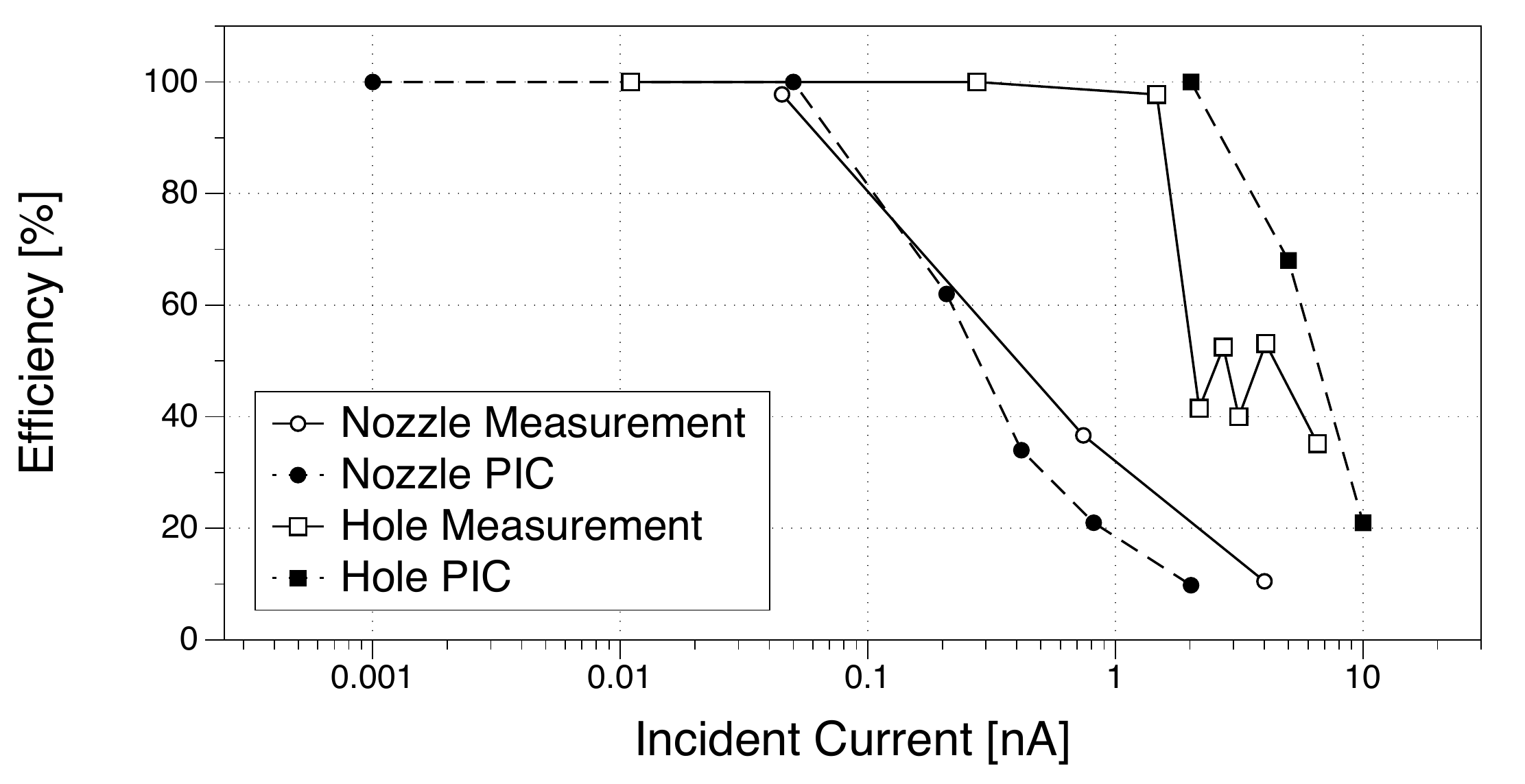}}
}%
\caption{Results from measurements and simulations of $\epsilon_{extraction}$ using a beam of $^{nat}$K$^+$ for two different 
orifices.  Measurements were performed at room temperature with a helium gas pressure of P~=~50~torr, RF amplitude of 
200~V$_{pp}$, and push field strength of E$_p$~=~5~V/cm.  Lines are to guide the eye.}
\label{rt_orifice_plot}
\end{figure}

\begin{figure*}[hbt!]
\center{
\resizebox{0.9\textwidth}{!}{%
\includegraphics{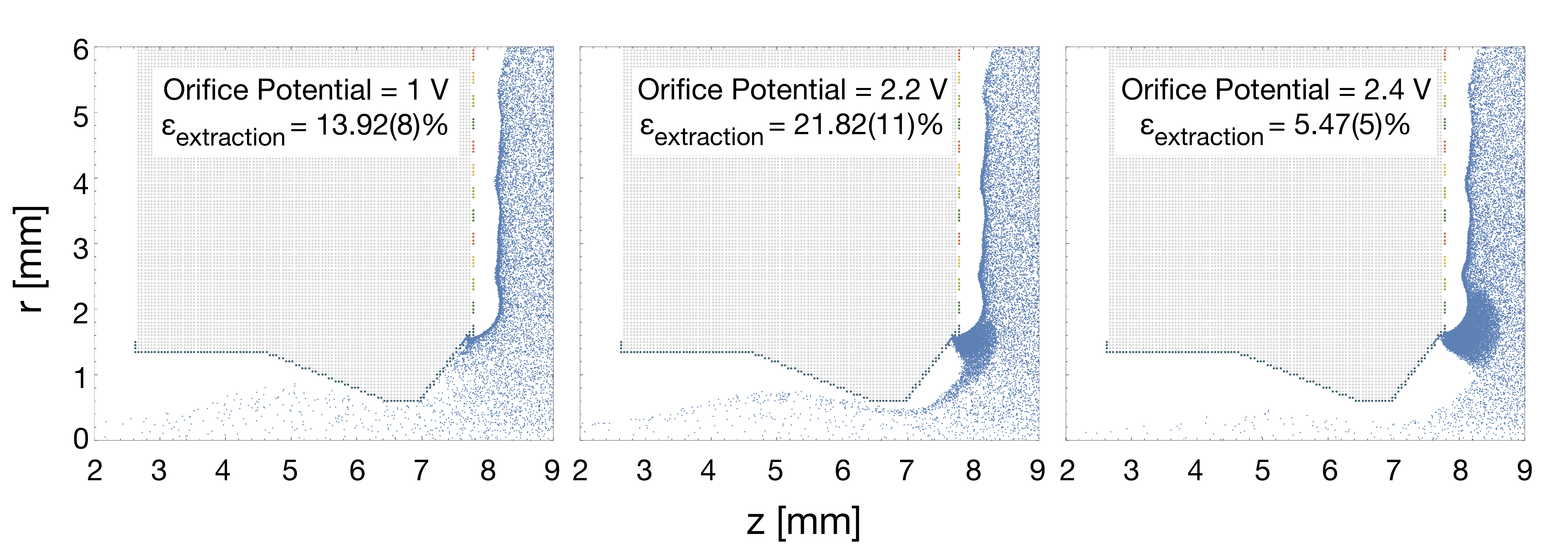}}
}%
\caption{Ion positions at t~=~3~ms during PIC simulation of a beam of O$_2^+$ with an intensity of 817~pA for three different 
potentials applied to the orifice.}
\label{nozzle_ions}
\end{figure*}

To validate the code, simulations were run and compared against this experimental data.  Preliminary 
simulation results showed that the original ACGS orifice, shown schematically in Fig.~\ref{mini_carpet_PIC_model}, 
might have lower ion throughput than a simple hole.  In order to test this, another mini RF carpet was fabricated 
with a simple hole as an orifice with a diameter of 1 mm and length of 0.8 mm.  The $\epsilon_{extraction}$ 
measurements were performed at room temperature with a helium gas pressure of P~=~50~torr for both orifices, 
E$_p$~=~5~V/cm, and the results are shown in Fig.~\ref{rt_orifice_plot}.  The agreement between simulation and 
experiment is good, and shows that there is a difference in the space charge capacity between the two orifice 
configurations, with the simple hole geometry increasing the throughput by about an order of magnitude and decreasing 
the sensitivity of $\epsilon_{extraction}$ on the potential applied to the orifice.  The reason is illustrated in 
Fig.~\ref{nozzle_ions}, where a snapshot of the ions' positions are shown together with the mini RF carpet and original 
orifice.  The initial current of O$_2^+$ ions is 817~pA, and the difference between the three plots is only the potential 
applied to the orifice (1~V, 2.2~V, and 2.4~V, respectively).  When the potential applied to the orifice is too low, 
such as in the 1~V case, the ions are not sufficiently repelled from the opening of the orifice.  When 2.2~V are applied 
to the orifice, which corresponds to maximum $\epsilon_{extraction}$, there are two noteworthy features.  
The first is that a local trap is formed near the edge of the mini RF carpet that fills up with ions.  The second is that 
the ions are more efficiently repelled from the throat of the orifice.  Increasing the potential applied to the orifice 
up to 2.4~V significantly reduces $\epsilon_{extraction}$ and increases the number of ions in the trapped region.  It's 
clear that the initial opening angle of the original orifice is responsible for the reduced performance when compared to 
the simple hole geometry, resulting in a very narrow range for tuning the orifice potential and vastly reducing 
the maximum $\epsilon_{extraction}$ achievable in the presence of space charge.  

\subsubsection{Extraction efficiency of simple hole geometry}
\label{hole_ext_sec}

\begin{figure*}[hbt!]
\center{
\resizebox{1\textwidth}{!}{%
\includegraphics{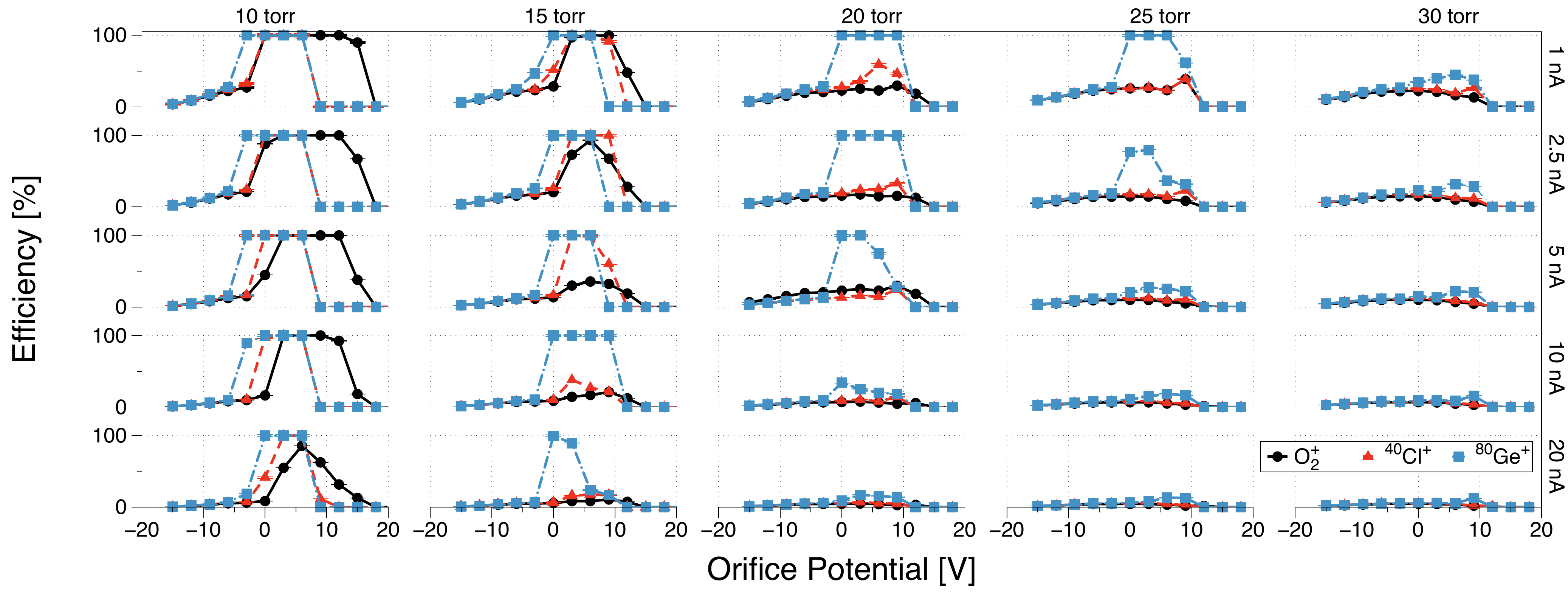}}
}%
\caption{Results of simulations of $\epsilon_{extraction}$ of O$_2^+$, $^{40}$Cl$^+$, and $^{80}$Ge$^+$ ions through the simple 
hole extraction geometry for various helium buffer gas pressures at T~=~50~K (columns) and incident O$_2^+$ ion currents (rows). 
An RF amplitude of 200~V$_{pp}$ on the mini RF carpet with an RF frequency of 6.6 MHz and applied push electric field of 
E$_p$~=~5~V/cm was used.}
\label{hole_extract_scan}
\end{figure*}

In Sec.~\ref{pic_vs_exp_sec} only the extraction efficiency of the O$_2^+$ ions were considered, but we are 
ultimately interested in the extraction efficiency of the isotopes of interest.  The significant amount of O$_2^+$ 
ions simultaneously being transported across the carpet and through the orifice will certainly impact $\epsilon_{extraction}$ of 
the isotopes of interest.  This effect was probed by running the simulation until the collection efficiency of the O$_2^+$ ions 
reached 100\% and then releasing 10,000 ions of both $^{40}$Cl$^+$ and $^{80}$Ge$^+$ and recording the fraction that was successfully 
transported through the orifice.  Compared to the lowest O$_2^+$ ion current investigated, 1~nA, the space charge effects generated 
by any realistic number of rare isotope ions is orders of magnitude lower and therefore not considered.  The reduced mobility for 
both $^{40}$Cl$^+$ and $^{80}$Ge$^+$ in a helium buffer gas were estimated to be 20~cm$^2$/(V$\cdot$s), and the results are not 
sensitive within the level of uncertainty of $\approx$10\%.  Fig.~\ref{hole_extract_scan} shows the results of $\epsilon_{extraction}$ 
for the three ion species, O$_2^+$, $^{40}$Cl$^+$, and $^{80}$Ge$^+$, as a function of the potential applied to the orifice for multiple 
buffer gas pressures (columns) and incident O$_2^+$ ion currents (rows).  An RF amplitude of V$_{RF}$~=~200~V$_{pp}$ was used on the 
mini RF carpet at a frequency of $\nu_{RF}$~=~6.6~MHz with a push field magnitude of E$_p$~=~5~V/cm.  In general, for low buffer gas 
pressure and low O$_2^+$ ion current 100\% extraction efficiency is obtained.  Increasing the buffer gas pressure and incident 
O$_2^+$ ion current results in a significant drop in $\epsilon_{extraction}$ for all ion species, but not equally, as the lighter 
ion species display a faster fall off in $\epsilon_{extraction}$.  For our set of parameters, 
V$_{eff,d}$($^{40}$Cl)~$\approx$~1.25$\cdot$V$_{eff,d}$(O$_2^+$) and V$_{eff,d}$($^{80}$Ge)~$\approx$~2.5$\cdot$V$_{eff,d}$(O$_2^+$), 
so the heavier ions experience a stronger repelling force from the mini RF carpet and are therefore more tolerant to the additional 
electric field due to space charge.  It should be noted that the range and widths of the efficiency curves vary with ion species, 
pressure, and incident O$_2^+$ ion current, and that the composition of the extracted beam could be tuned to some degree by varying 
the orifice potential. 

\begin{figure}[h]
\center{
\resizebox{0.5\textwidth}{!}{%
\includegraphics{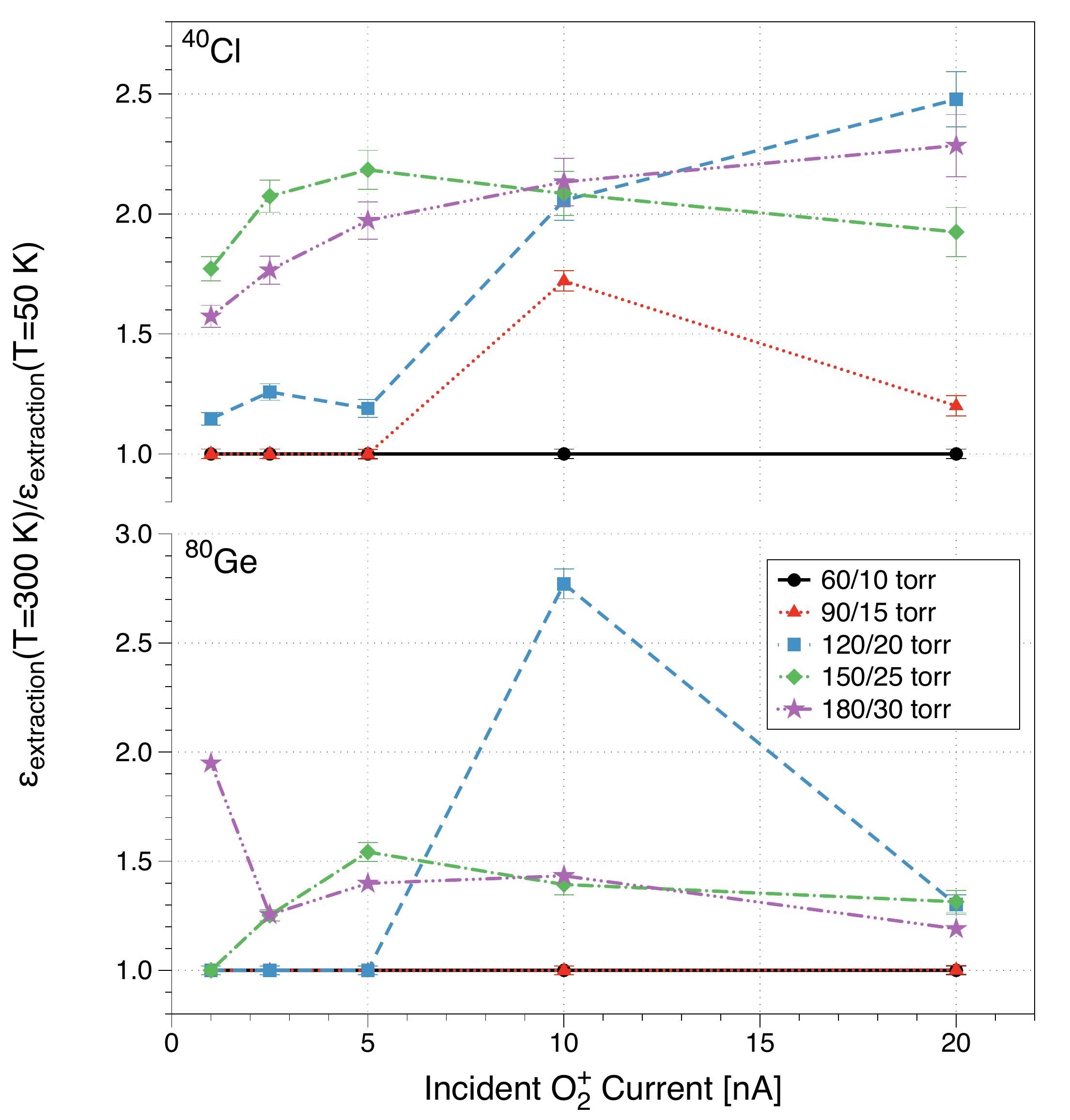}}
}%
\caption{Ratios of the maximum values of $\epsilon_{extraction}$ at T~=~300~K and T~=~50~K for $^{40}$Cl (top) and $^{80}$Ge (bottom).  
The buffer gas pressures were chosen such that the volume density remains constant for the two temperature values.  Line are to guide 
the eye.}
\label{acgs_orifice_ratios}
\end{figure}

Finally, the impact on $\epsilon_{extraction}$ of cooling the helium buffer gas to cryogenic temperatures was explored.  It might be 
expected that if the volume density of the buffer gas is constant, there should not be an impact on $\epsilon_{extraction}$, 
as should also be the case with $\epsilon_{stopping}$ and $\epsilon_{transport}$.  However, since the pressure differential across the 
orifice is reduced in the case of a cryogenic buffer gas, this leads to reduced mass flow and gas velocities. Since the gas flow 
in the region of the orifice does play a role in the extraction of ions out of the ACGS, it is worth investigating.  The same 
simulations shown in Fig.~\ref{hole_extract_scan} were run with a buffer gas temperature of T~=~300~K at pressures corresponding to the 
same volume densities.  The results are shown in Fig.~\ref{acgs_orifice_ratios}, where the ratios of the maximum values of 
$\epsilon_{extraction}$ for the same buffer gas volume densities at the two temperatures, T~=~300~K and 50~K, are plotted for both 
$^{40}$Cl and $^{80}$Ge with respect to the incident O$_2^+$ current.  For low volume densities and incident O$_2^+$ currents, the ratio 
is one, as $\epsilon_{extraction}\approx100\%$ is achieved for both buffer gas temperatures.  In the case of $^{40}$Cl, looking at 
Fig.~\ref{hole_extract_scan} shows the lowest volume density where the maximum achievable $\epsilon_{extraction}$ falls below 100\% 
is at P~=~15~torr, at T~=~50~K, with an incident O$_2^+$ current of 10 nA.  Looking at the corresponding data in 
Fig.~\ref{acgs_orifice_ratios} shows that the ratio jumps up from 1 to approximately 1.7, reflecting a greater reduction in 
$\epsilon_{extraction}$ for the lower buffer gas temperature.  The same effect is observed in $^{80}$Ge, also for an incident 
O$_2^+$ current of 10 nA, but for case of P~=~20~torr at T~=~50~K.  In fact, the higher-temperature buffer gas always yields a better 
$\epsilon_{extraction}$ as long as the maximum achievable $\epsilon_{extraction}$ has fallen below 100\% for the cryogenic buffer 
gas case, and generally has a greater impact on the lower-mass $^{40}$Cl.

\section{Summary and Conclusions}
A framework for the simulation of space-charge effects in gas cells was developed and applied to the study of FRIB/NSCL's newest 
linear gas cell, ACGS.  In particular, the impact of He$^+$/e$^-$ pairs generated during the stopping process on $\epsilon_{transport}$, 
and the impact on $\epsilon_{extraction}$ due to large O$_2^+$ ion currents, were studied.  The simulations were benchmarked against 
experimental results, where available, and good agreement was obtained.  A bottle neck in the extraction system was identified via 
simulation and corrected.  The simulations showed that high $\epsilon_{stopping}$ and $\epsilon_{transport}$ are achievable up to 
10$^8$ pps for medium-mass isotopes with A/Q~$\approx$~80, e.g., $^{80}$Ge$^+$.  The performance for lighter-mass isotopes, around 
A/Q~$\approx$~40, e.g., $^{40}$Cl, starts to decrease after 10$^6$ pps.  In both cases, the buffer gas pressure and strength of the 
applied push field, E$_p$, must be varied based on the incident ion rate in order to achieve optimal efficiency.  Moreover, it was 
found that a large stable-beam current, generated by charge exchange with He$^+$ ions, can significantly decrease 
$\epsilon_{extraction}$, particularly for ions with with A/Q~$\le$~40.  Finally, it was shown that using a helium buffer gas at 
cryogenic temperatures can negatively impact $\epsilon_{extraction}$ when operating at the higher volume densities and incident 
O$_2^+$ currents considered in this paper.

Key goals in a next-generation linear gas cell to maintain good efficiency for high injected ion rates would be faster He$^+$/e$^-$
removal and reduced stable ion density near the extraction orifice.  Increasing the length of the stopping chamber would allow 
for lower buffer gas pressures which increases the He$^+$/e$^-$ collection rate, increases push-field tolerance of ions on the 
main RF carpet, and increases the ion transport speed.  In addition, adding multiple layers of RF carpets would reduce the 
distance He$^+$/e$^-$ pairs need to travel, increasing collection speed.  Multiple layers of RF carpets would also make guiding 
the ions to multiple extraction points easier, thereby reducing the ion density near the orifice and compensating for reductions 
in $\epsilon_{extraction}$ due to cryogenic operation.

\section*{Acknowledgements}
This work was conducted with the support of Michigan State University, the National Science Foundation under Contract No. 
PHY-1565546, and the Department of Energy under Grant No.~DE-SC0021423. Computational resources and services were provided 
by the Institute for Cyber-Enabled Research at Michigan State University.




\bibliographystyle{elsarticle-num} 
\bibliography{subset.bib}





\end{document}